\documentstyle[12pt,apjpt4,tighten,amstex]{article}
\input epsf
\def\eps@scaling{.95}
\def\epsscale#1{\gdef\eps@scaling{#1}}
\def\plotone#1{\centering \leavevmode
\epsfxsize=\eps@scaling\columnwidth \epsfbox{#1}}
\def\plottwo#1#2{\centering \leavevmode
\epsfxsize=.49\columnwidth \epsfbox{#1} \hfil
\epsfxsize=.49\columnwidth \epsfbox{#2}}
\begin{document}
\def\czlg{cz_{{\rm LG}}}
\def\uprime{u^\prime}
\def\udoubleprime{u^{\prime\prime}}
\def\gtsima{$\, \buildrel > \over \sim \,$}
\def\ltsima{$\, \buildrel < \over \sim \,$}
\def\simgt{\lower.5ex\hbox{\gtsima}}
\def\simlt{\lower.5ex\hbox{\ltsima}}

\def\sm{$\sim\,$}
\def\smgt{$\simgt\,$}
\def\smlt{$\simlt\,$}
\def\smeq{$\simeq\,$}
\def\onesigma{$1\,\sigma$}
\def\nhat{\ifmmode {\hat{\bf n}}\else${\hat {\bf n}}$\fi}

\def\degs{\ifmmode^\circ\,\else$^\circ\,$\fi}
\def\kps{\ifmmode{\rm km}\,{\rm s}^{-1}\else km$\,$s$^{-1}$\fi}
\def\kms{\ifmmode{\rm km}\,{\rm s}^{-1}\else km$\,$s$^{-1}$\fi}
\def\ksmpc{\ifmmode{\rm km}\,{\rm s}^{-1}\,{\rm Mpc}^{-1}\else km$\,$s$^{-1}\,$Mpc$^{-1}$\fi}
\def\kmsmpc{\kms\ {{\rm Mpc}}^{-1}}
\def\etal{{\sl et al.}}
\def\ie{{\it i.e.}}
\def\eg{{\it e.g.}}
\def\apriori{{\rm a priori}}
\def\aposteriori{{\rm a posteriori}}
\def\halpha{H$\alpha$}
\def\h1{$h^{-1}$}
\def\dnsigma{$D_n$-$\sigma$}
\font\tensm=cmcsc10
\def\hii{H\kern 2.0pt{\tensm ii}}	
\def\hi{\ifmmode{\rm H\kern 2.0pt{\tensm I}}\else H\kern 2.0pt{\tensm I}\fi}	
\def\potent{{\small POTENT}}
\def\potiras{{\small POTIRAS}}
\def\simpot{{\small SIMPOT}}

\def\dinv{d_{{\rm inv}}}
\def\deltainv{\Delta_{{\rm inv}}}
\def\muinv{\mu(\dinv)}
\def\delinv{\Delta_{{\rm inv}}}
\def\onehalf{\frac{1}{2}}
\def\sigz{\sigma_z}
\def\sigeta{\sigma_{\eta}}
\def\sigr{\sigma_r}
\def\sigv{\sigma_v}
\def\sigin{\sigma_{{\rm in}}}
\def\sigout{\sigma_{{\rm out}}}
\def\sigM{\sigma_M}
\def\sigrb{\sigma_{\rm RB}}
\def\etazero{\eta^0}
\def\etapr{\eta^{0\prime}}
\def\nsigeta{\frac{1}{\sqrt{2\pi}\,\sigeta}}
\def\nsigr{\frac{1}{\sqrt{2\pi}\,\sigr}}
\def\nsigv{\frac{1}{\sqrt{2\pi}\,\sigv}}
\def\nsigM{\frac{1}{\sqrt{2\pi}\,\sigM}}
\def\nsig{\frac{1}{\sqrt{2\pi}\,\sigma}}
\def\nsigD{\frac{1}{\sqrt{2\pi}\,\Delta}}
\def\nsigDinv{\frac{1}{\sqrt{2\pi}\,\delinv}}
\def\nsigzero{{1 \over \sqrt{2\pi}\,\sigma_0}}
\def\nsigm0{{1 \over \sqrt{2\pi}\,\sigma_{0m}}}
\def\nsigxi{\frac{1}{\sqrt{2\pi}\,\sigma_\xi}}
\def\nsigz{\frac{1}{\sqrt{2\pi}\,\sigz}}
\def\2overpi{2 \over \pi}
\def\expmM{\exp\!\left(-\frac{\left[m-(M(\eta)+\mu(r))\right]^2}{2\sigtf^2} \right)}
\def\expmMm0{\exp\!\left(-\frac{(m-(M(\eta_0)+\mu(r)))^2}{2\sigma_{0m}^2} \right)}
\def\expeta{e^{-\frac{(\eta-\eta_0)^2}{2\sigeta^2}}}
\def\expet{\exp\!\left(-\frac{\left[\eta - \eta^0(m-\mu(r))\right]^2}{2\sigeta^2}\right)}
\def\expetm{\exp\!\left(-\frac{[\eta - \eta^0(m-\mu)]^2}{2\sigeta^2}\right)}
\def\expmm{\exp\!\left(-\frac{(m-m(\eta))^2}{2\sigma^2}\right)}
\def\expmu{\exp\!\left(-\frac{(\mu(r) - \mu(m,\eta))^2}{2 \sigma^2} \right)}
\def\expmuD{\exp\!\left(-\frac{(\mu(r) - \mu(d))^2}{2 \sigma^2} \right)}
\def\explnrD{\exp\!\left(-\frac{\left[\ln r/d \right]^2}{2 \Delta^2} \right)}
\def\explnrDinv{\exp\!\left(-\frac{\left[\ln r/\dinv \right]^2}{2 \delinv^2} \right)}
\def\explnvD{e^{-\frac{\left[\ln\frac{v_c}{\vd}\right]^2}{2 \Delta^2}}}
\def\explnu{e^{-\frac{\left[\ln\frac{v_r-u}{v_c}\right]^2}{2 \Delta^2}}}
\def\explnx{e^{-\frac{\left(\ln x\right)^2}{2 \Delta^2}}}
\def\explnxinv{e^{-\frac{\left(\ln x\right)^2}{2 \delinv^2}}}
\def\expvr{\exp\!\left(-\frac{[v_r - (v_c+ v_p(v_c))]^2}{2\sigr^2}\right)}
\def\expcz{\exp\!\left(-\frac{[cz - (r+ u(r))]^2}{2\sigv^2}\right)}
\def\expxi{\exp\!\left(-\frac{(\xi - \xi(m,\eta))^2}{2 \sigma_{\xi}^2} \right)}
\def\expz{\exp\!\left(-\frac{(m_z-m_z(m,\eta))^2}{2\sigz^2} \right)}
\def\calb{{\cal B}}
\def\cali{{\cal I}}
\def\calbinv{{\cal B}_{\rm inv}}
\def\cala{{\cal A}}
\def\xlim{\xi_\ell}
\def\xint{\int_{\xlim}^\infty\,}
\def\xxint{\int^{\xlim}_{-\infty}\,}
\def\erf{{\rm erf}}
\def\infint{\int_{-\infty}^\infty}
\def\inf0int{\int_0^\infty}
\def\mlim{m_\ell}
\def\mlint{\int_{-\infty}^{\mlim}\,}
\def\Aint{\int_{-\infty}^\cala\,}
\def\onplserf{\left[1+\erf\left(\frac{\xi(m,\eta)-\xlim}{\sqrt{2}\,\sigma_\xi}\right)\right]}
\def\oneplserf{\left[1+\erf\left(\frac{\xi(m,\eta^0(m-\mu(r)))-\xlim}{\sqrt{2}\,\sigma_\xi}\right)\right]}
\def\sigxi{\sigma_\xi}
\def\Axi{\cala_{\xi}}
\def\Axiinv{\cala_{\xi,{\rm inv}}}
\def\Az{\cala_z}
\def\Azinv{\cala_{z,{\rm inv}}}
\def\Axiprime{\cala'_{\xi}}
\def\Azprime{\cala'_z}
\def\betafac{\frac{\beta}{\sqrt{1+\beta^2}}}
\def\alfafac{\frac{\alpha}{\sqrt{1+\alpha^2}}}
\def\Abetafac{\frac{\Axi^2}{1+\beta^2}}
\def\Abfsqrt{\frac{\Axi}{\sqrt{1+\beta^2}}}
\def\Aalfafac{\frac{A_z^2}{1+\alpha^2}}
\def\Aafsqrt{\frac{A_z}{\sqrt{1+\alpha^2}}}
\def\atan{{\rm tan}^{-1}\,}
\def\mij{m_{ij}}
\def\etaij{\eta_{ij}}	
\def\dij{d_{ij}}
\def\vij{v_{ij}}
\def\muij{\mu_{ij}}

\def\dugc{\ifmmode D_{{\rm UGC}}\else$D_{{\rm UGC}}$\fi}
\def\logdugc{\ifmmode \log\dugc\else$\log\dugc$\fi}
\def\deso{\ifmmode D_{{\rm ESO}}\else$D_{{\rm ESO}}$\fi}
\def\logdeso{\ifmmode \log\deso\else$\log\deso$\fi}
\def\iras{{\sl IRAS\/}}
\def\potent{{\small POTENT\/}}
\def\itf{{\small ITF\/}}
\def\velmod{{\rm VELMOD\/}}
\def\bfv{{\bf v}}
\def\verr{\bfv_{{\rm err}}}
\def\bfw{{\bf w}}
\def\bfwlg{\bfw_{{\rm LG}}}
\def\bfV{{\bf V}}
\def\bfr{{\bf r}}
\def\bfs{{\bf s}}
\def\bfq{{\bf q}}
\def\bfu{{\bf u}}
\def\bfx{{\bf x}}
\def\bfk{{\bf k}}
\def\bfz{{\bf z}}
\def\bfn{{\bf n}}
\def\sigtf{\sigma_{{\rm TF}}}
\def\vev#1{{\left\langle#1\right\rangle}}
\def\like{{\cal L}}
\def\likeforw{\like_{{\rm forw}}}
\def\likeinv{\like_{{\rm inv}}}

\title{ Maximum-Likelihood Comparison of Tully-Fisher \\
and Redshift Data. II. Results from an Expanded Sample}
\author{Jeffrey A.\ Willick$^a$ and Michael A.\ Strauss$^{b,c,d}$}
\bigskip
{\small\centerline{
$^a$ Dept.\ of Physics, Stanford University,
Stanford, CA 94305-4060
{\tt (jeffw@@perseus.stanford.edu)}} 
\centerline{$^b$ Princeton University Observatory, Princeton University,
Princeton, NJ 08544
{\tt (strauss@@astro.princeton.edu)}}
\centerline{$^c$ Alfred P.\ Sloan Foundation Fellow}
\centerline{$^d$ Cottrell Scholar of Research Corporation}}
\begin{abstract}
This is the second in a series of papers in which we compare Tully-Fisher (TF)
data from the Mark III Catalog with predicted peculiar
velocities based on the \iras\ galaxy redshift survey and gravitational
instability theory, using a rigorous
maximum likelihood method called \velmod. In Paper I (Willick
\etal\ 1997b), we applied the method
to a $cz_{\rm LG} \leq 3000\ \kms,$ 838-galaxy TF sample
and found $\beta_I=0.49\pm 0.07,$ where $\beta_I\equiv
\Omega^{0.6}/b_I$ and $b_I$ is the linear biasing parameter for
\iras\ galaxies. In this paper 
we increase the redshift limit to $cz_{\rm LG}=7500\ \kms,$ thereby enlarging
the sample to 1876 galaxies.  The expanded sample
now includes the % Willick (1991; W91PP) Perseus-Pisces and Courteau-Faber 
% (Courteau 1995; CF) 
W91PP and CF subsamples of the Mark III catalog, in addition
to the A82 and MAT
subsamples already considered in Paper I. 

We implement \velmod\ using
both the forward and inverse forms of the TF relation,
and allow for a more general form of the quadrupole
velocity residual detected in Paper I.
We find $\beta_I=0.50\pm 0.04$ (\onesigma\ error) at 300 \kms\
smoothing of the \iras-predicted velocity field. 
The fit residuals are spatially
incoherent for $\beta_I=0.5,$ indicating
that the \iras\ plus quadrupole velocity field is a good fit
to the TF data.
%in all cases in which
%we allow for the quadrupole velocity residual. 
%Our best-fit quadrupole closely resembles the Paper I quadrupole
%at small distances, but cuts off rapidly beyond
%\sm 3500 \kms. 
If we eliminate the quadrupole 
we obtain a worse
fit, but a similar value for $\beta_I$ of $0.54\pm 0.04.$
%The above results are quoted for $300\ \kms$ smoothing
%of the \iras\ density field. 
Changing the \iras\ smoothing
scale to $500\ \kms$ has almost no effect on the best $\beta_I.$
We find 
%moderately strong 
evidence for a density-dependence
of the small-scale velocity dispersion,
%. This dependence is well modeled by 
$\sigma_v(\delta_g)\simeq (100 + 35\,\delta_g)\ \kms.$

%\velmod\ yields TF calibrations along with the maximum likelihood
%solution for $\beta_I.$ 
We confirm our Paper I result that the %\velmod\ 
TF relations for the A82 and MAT samples found by \velmod\  
are consistent with those that went into the published Mark III catalog. 
However, the \velmod\
TF calibrations for the W91PP and CF samples 
place objects \sm 8\% closer
than their Mark III catalog distances, which has an important effect
on the inferred large-scale flow field at 4000--6000 \kms.  With this
recalibration, the \iras\ and Mark III velocity fields are consistent
with one another at all radii. 
%This suggests that analyses
%requiring precalibrated TF relations, such as the POTIRAS approach
%(Sigad \etal\ 1997), may require modification.
%Observational programs aimed at recalibrating the Mark III TF relations 
%independently of \iras\ should clarify this issue in the near future.
\end{abstract}

\section{Introduction}
\label{sec:intro}
In recent years, a number of groups have compared
the peculiar velocity and/or density fields derived
from distance indicator data with the corresponding fields obtained
from redshift survey data (Kaiser \etal\ 1991; Dekel
\etal\ 1993; Hudson 1994; Roth 1994; Hudson \etal\ 1995; Schlegel 1995; Davis, Nusser, \& Willick
1996; Willick \etal\ 1997b, hereafter Paper I; 
da Costa \etal\ 1997; Riess \etal\ 1997; Sigad \etal\ 1998). The principal
goals of these comparisons are to test the gravitational instability
(GI) picture for the growth of large-scale structure and to
measure the 
parameter $\beta=\Omega^{0.6}/b,$ where $\Omega$ is the
present value of the cosmological density parameter, and $b$
is the ``biasing parameter'' (see below).
A longer-range goal is to measure $\Omega$ itself, by combining
the $\beta$-measurement with 
other measurements that constrain a
combination of $\Omega$ and $b.$ 

Measurement of $\beta$ is based on the relationship
between the peculiar velocity and density fields
predicted by GI for the linear regime (Peebles 1980):
\begin{equation}
\bfv_p(\bfr) = \frac{\beta}{4\pi} \int d^3\bfr'\,\frac{\delta_g(\bfr')
(\bfr' - \bfr)}{\left|\bfr' - \bfr\right|^3} \,.
\label{eq:vpdelta}
\end{equation}
In equation~(\ref{eq:vpdelta}), the galaxy number
density fluctuation field $\delta_g$ is assumed to be related to the underlying mass
density fluctuation field $\delta$ by the simple linear biasing model $\delta_g=b\,\delta.$ 
Taking the divergence of both sides of equation~(\ref{eq:vpdelta})
yields:
\begin{equation}
\nabla \cdot \bfv_p = -\beta\,\delta_g\,.
\label{eq:div_v}
\end{equation}
In both equations, distances are assumed to be measured in units
of the Hubble velocity (i.e. $H_0\equiv 1$).

\def\vv{$v$-$v$}
\def\dd{$d$-$d$}
To estimate the quantity $\beta$ via equation~(\ref{eq:vpdelta}),
one measures $\delta_g$ from redshift survey data, and then {\em predicts\/} $\bfv_p(\bfr)$
for a sample of
galaxies with redshift-independent distances\footnote{Hereafter
we will assume for definiteness that the redshift-independent
distances have been derived from the Tully-Fisher (1977;
TF) relation. 
However, the method
applies to any distance indicator relation.} 
and thus estimated
peculiar velocities.  
One then asks, for what value of $\beta$ does the velocity
field prediction best fit the TF data?
This approach is known as the ``velocity-velocity'' (\vv) comparison. 

Alternatively, one can do a 
``density-density'' (\dd) comparison,
using equation~(\ref{eq:div_v}). 
In this case the crucial input from the redshift survey is not
the predicted velocity field $\bfv_p(\bfr),$ 
but instead the directly observed density field $\delta_g.$
However, the TF data must now be converted into a three-dimensional
velocity field, whose divergence is then taken to yield an effective
mass density field $-\nabla\cdot\bfv_p.$ Comparison of
$\delta_g$ and $-\nabla\cdot\bfv_p$, via equation~(\ref{eq:div_v}), then yields $\beta.$

In the \vv\ comparison the redshift survey data is 
manipulated to yield predicted peculiar
velocities (see, e.g., Yahil \etal\ 1991). 
The way this
predicted velocity field changes with $\beta$ is what
provides the \vv\ comparison with its discriminatory power.
In the \dd\ comparison 
it is the numerical processing of the TF data which is more important for $\beta$-determination.
This is done using the 
POTENT method and its variants (Bertschinger \& Dekel 1989; 
Dekel, Bertschinger, \& Faber 1990; Dekel 1994, 1997;
da Costa \etal\ 1996)
which invoke the assumption of potential flow in order to convert
the radial
TF data into a 3-dimensional velocity field,
and thus into an effective mass density field. 

The redshift survey most often used in recent
\vv\ and \dd\ comparisons, and the one we use in this paper, is the
1.2 Jy \iras\ redshift 
survey (Fisher \etal\ 1995), which covers nearly the full sky 
and is only weakly affected by dust extinction and related
effects at low Galactic latitude. 
Hereafter, we write $b_I$ to denote the \iras\ biasing parameter,
and $\beta_I=\Omega^{0.6}/b_I.$ 

The published results for $\beta_I$ appear to bifurcate 
according to
whether the \dd\ or the \vv\ comparison is used. The former has
been implemented using the POTENT method % of the density field
by Dekel \etal\ (1993), Hudson \etal\ (1995), and Sigad \etal\ (1998; hereafter POTIRAS)
to obtain $\beta_I=1.29\pm 0.30,$ $\beta_I\simeq 1.0\pm 0.13,$ and $\beta_I=0.89 \pm 0.12,$
respectively\footnote{The Hudson \etal\ result has been converted from
the measured $\beta_{opt}$ assuming that $b_{opt}/b_I = 1.3$
(Strauss \etal\ 1992).}
(the error bars are \onesigma). 
In the first of these studies, POTENT was applied to the redshift-independent
distances in the Mark II Catalog (Burstein 1989), while in the latter two it was
applied to those in the Mark III Catalog (Willick \etal\ 1997a).
These relatively high values of $\beta_I$ have often been
cited (assuming the that $b_I$ is not much different from unity)
as evidence for an $\Omega=1$ universe.
In contrast, the \vv\ approach 
has typically produced lower values of
$\beta_I,$ 
which (again assuming that $b_I\approx 1$) point to
a low-density ($\Omega\simeq 0.2$--$0.5$) universe.
Davis, Nusser, \& Willick (1996)
and da Costa \etal\ (1997) each found $\beta_I=0.6\pm 0.15$ by
applying the inverse Tully-Fisher (ITF)
method of Nusser \& Davis (1995), 
in the former case to the Mark III 
catalog and in the latter case to the SFI sample of Giovanelli \etal\
(1997). 
Riess \etal\ (1997) also used the ITF method for
distances obtained from Type Ia supernovae, 
finding $\beta_I=0.40\pm 0.15.$ Roth (1994) and Schlegel (1995) used
\vv\ analyses of smaller TF samples to obtain $\beta_I=0.6$ and $\beta_I=0.39$ respectively.
Shaya, Peebles, \& Tully (1995) find $\beta_{I} = 0.45\pm 0.15$
from their \vv\ analysis of nearby TF data\footnote{We have 
again converted their $\beta_{opt}$ into an equivalent $\beta_I$ assuming
$b_{opt}/b_I=1.3.$}. 
Finally, we found $\beta_I=0.49\pm 0.07$ in Paper I by applying 
the \velmod\ method (\S~2)
to a subset of the Mark III Catalog restricted to $\czlg\leq 3000\
\kms.$ 
See Strauss \& Willick (1995, hereafter SW) for a review of these and other
methods for measuring $\beta$. 

In this paper we will again apply \velmod, now to an
expanded sample 
that includes all Mark III Catalog field spirals out to $\czlg=7500\ \kms.$
This larger sample will lead to tighter constraints on $\beta_I$
than obtained in Paper I, although our results will be fully consistent.
The outline
of this paper is as follows. In \S~\ref{sec:method} we review the \velmod\ method.
In \S~\ref{sec:select} we describe the selection of our expanded sample. 
In \S~\ref{sec:quad}, we discuss the motivation behind and implementation
of a more general form of the quadrupole velocity residual
introduced in Paper I. In \S~\ref{sec:results}, we present the main
results of the maximum likelihood analysis.  
In \S~\ref{sec:calib}, we compare the \velmod\ TF calibrations to
those in Mark III.  In \S~\ref{sec:residuals}, we quantify the
goodness of fit of our model to the data.  Finally, in
\S~\ref{sec:summary} we summarize our main conclusions. In the
Appendix, we describe an analytic approximation to computing the
\velmod\ likelihoods.  The formulae presented there are not limited to
\velmod\ and are generally useful in velocity field analyses.

\section{Method of Analysis}
\label{sec:method}
\subsection{The VELMOD Approach}
\label{sec:velmod-approach}
\velmod\ is a maximum likelihood method
for comparing TF data to
predicted peculiar velocity fields. 
The method was
described in some detail in Paper I, \S 2, and we give only a brief
overview here. The TF data for each galaxy consist of its direction
$(l,b)$, its redshift $cz$ measured in the Local Group (LG) frame, 
its apparent magnitude $m$, and its velocity width parameter
$\eta \equiv
\log (\Delta v) - 2.5.$
The velocity field model gives the
relationship between redshift and distance ($r$) along any given line of
sight, albeit with some finite scatter, $\sigma_v,$ due to inaccuracies
of the model and small-scale velocity ``noise.''  
We assume that there exists a forward 
[$M(\eta)$] and an inverse [$\eta^0(M)$] 
TF relation for
each sample, such that $m,$ $\eta,$ and $r$ are related as follows:
\begin{equation} 
m = M(\eta)+5\log r  = A - b\,\eta + 5\log r\,,\ \hbox{with rms dispersion
$\sigtf$}
\label{eq:TF-forward}
\end{equation}
(forward relation), or
\begin{equation} 
\eta = \eta^0(m-5\log r) = - e\left(m-5\log r - D\right),\ \hbox{with rms dispersion $\sigma_{\eta}$}
\label{eq:TF-inverse} 
\end{equation}
(inverse relation). We refer to $A,$ $b,$ and $\sigtf$ 
($D,$ $e,$ and $\sigeta$) as the zero point, slope, and scatter
of the forward (inverse) TF relation, or simply as the TF parameters.

For each object in the TF sample, 
$P(m|\eta,cz)$---the probability that a galaxy of redshift
$cz$ and velocity width parameter $\eta$ will have apparent magnitude
$m$---is evaluated when the forward TF relation is used; see
equation~(\ref{eq:pmgetaz}).
For the inverse TF relation,
it is $P(\eta|m,cz)$ that is evaluated, equation~(\ref{eq:petagmz}).
These single-object probabilities depend on a number of parameters: 
\begin{enumerate}
\item The three TF parameters for each distinct subsample
(i.e., A82, MAT, W91PP, and CF).
In \S~\ref{sec:TFscatter}, we explore the addition of a fourth TF parameter
describing the change in the scatter with luminosity.  
\item $\beta_I,$ which determines the \iras-predicted peculiar velocity. 
\item The small-scale velocity dispersion $\sigma_v.$ In \S~\ref{sec:sigmav},
we include an additional parameter $f_\delta$ describing the density
dependence of $\sigma_v$. 
\item A cutoff scale, $R_Q,$ for the velocity quadrupole (\S~\ref{sec:quad}).
\item A LG velocity vector $\bfwlg,$ required because 
small errors in the prediction of the LG velocity propagate to all other 
peculiar velocity predictions (cf.\ Paper I, \S~2.2.3).
As $\bfwlg$ is primarily determined by nearby galaxies,
in this paper we simply fix it to its
Paper I value. 
\end{enumerate}

The single-object
probabilities are multiplied together, yielding an overall probability $P$
for the entire TF sample. The value of $\beta_I$ for which
$P$ is maximized 
is the maximum likelihood value of $\beta_I.$ 
In practice, rather than maximizing $P$ we minimize 
$\like\equiv -2\ln P.$
(In \S~5, we will write
$\likeforw$ and $\likeinv$ to distinguish forward
from inverse likelihoods.)
A single \velmod\ run consists of minimizing $\like,$
at each of 10 values of $\beta_I,$ $0.1,0.2,\ldots,1.0,$
by continuously varying the TF parameters of each
sample\footnote{We hold the velocity parameters
$\sigv$ and $R_Q$ fixed in any given \velmod\ run, but carry out
a series of runs in which they take on a range of discrete values,
and in this way determine their maximum likelihood values;
cf.\ \S~5.3 and 5.4.
The only velocity parameter treated as
continuously variable is $f_\delta.$}. 
A cubic fit to the $\like(\beta_I)$ points then yields the
maximum-likelihood value of $\beta_I.$ 
Tests with mock catalogs,
discussed in Paper I, demonstrated that this maximum likelihood value of 
$\beta_I$ is an unbiased estimator of the true value when the \iras\
peculiar velocities are predicted using a 300 \kms\ Gaussian smoothing scale
and a Wiener filter. The tests also showed that rigorous
\onesigma\ errors in $\beta_I$ are given by noting
the values at which $\like$ differs by one unit from its minimum value,
as obtained from the cubic fit.

Because the TF parameters for each sample
are determined via maximum likelihood,
\apriori\ TF calibrations are not required for \velmod.
Indeed, each value of $\beta_I$ is given the fairest possible
chance to fit the data by finding 
the TF parameters most in accord with the velocity field
it produces.
These TF parameters are {\em not\/} constrained to be similar
to those used to produce the Mark III catalog distances 
(we discuss this issue further in \S~\ref{sec:calib}).
Furthermore, while the TF
scatter is treated as a free parameter, 
we emphasize that {\em maximizing likelihood is not
equivalent to minimizing scatter} (cf.\ Paper I, \S~3.4). 
In general, the minima of $\like$ and
of $\sigtf$ (or $\sigeta$) for a given subsample do not precisely coincide. 

\subsection{Implementation of Inverse \velmod}
\label{sec:inverse-velmod}
Because selection effects on the forward TF relation are strong
(Willick 1994), the sample selection function must be properly
modeled in forward \velmod\ in order to obtain unbiased results.
However, as 
selection depends weakly on velocity width, 
errors in modeling the selection function
will have little effect on inverse \velmod\ or comparable analyses. 
For this reason, inverse TF methods have been favored by
many workers (e.g., Schechter 1980; Aaronson \etal\ 1982b; Tully 1988;
Nusser \& Davis 1995; Shaya \etal\ 1995; da Costa \etal\ 1997; cf.\ SW
for a discussion). 
On the other hand, inverse, but not forward, \velmod\ 
depends on the
galaxy luminosity function $\Phi(M)$ 
(cf.\ Paper I, \S~2), 
which is not easy to quantify, given
the fact that each sample uses its own photometric system.  

Because $\Phi(M)$ appears in the integrals in both the
numerator and denominator of the expression for $P(\eta|m,cz)$
(equation~[\ref{eq:petagmz}]),  
it is not crucial to model it perfectly. 
We determine $\Phi(M)$ for each sample as follows.  As $\eta$ is
defined in essentially the same way for each sample, we assume
that there is a universal 
$\eta$-distribution function, $\phi(\eta),$ which we take to be a Gaussian
of mean $\eta_0=-0.05$ and dispersion $\Sigma_\eta=0.15.$ 
This distribution
function matches well what is seen in the Mark III TF samples 
above the cutoff $\eta_{{\rm min}} \sim -0.4$
imposed by magnitude and diameter limit effects. 
We then calculate $\Phi(M)$ 
using the relationship between $\phi(\eta)$ and 
$\Phi(M)$ given by
the TF relation itself:
\begin{equation}
\Phi(M) \approx \left|\phi\left[\eta^0(M)\right] \frac{d\eta^0}{d
M}\right| = e\,\phi\!\left[\eta^0(M)\right] \,,
\label{eq:Phiphi}
\end{equation}
where $\eta^0(M)$ is the inverse TF relation and $e$ is its slope 
(cf.\ equation~[\ref{eq:TF-inverse}]). 

The luminosity function obtained from equation~(\ref{eq:Phiphi})
is, as required, different for each sample, because each sample has its own TF
parameters.  
The differences reflect
bandpass effects and
differing approaches to extinction/inclination corrections for each
of the individual Mark III TF samples (Willick \etal\ 1997a). 
Ultimately, we will test the suitability of this
approximation by comparing the results of the
forward and inverse \velmod\ calculations.
To the extent they agree, we can be confident
that our imperfect modeling of the selection and
luminosity functions do not bias the results.

\subsection{An Analytic Approximation to the \velmod\ Likelihoods}
\label{sec:methoda}
A drawback of the original 
\velmod\ algorithm 
was its 
repeated evaluation of the 
numerical integrals in terms of which
the single-object likelihoods are defined. 
These integrations are crucial 
in triple-valued or flat zones in the redshift-distance relation
(cf.\ Paper I, \S~2.2.2).
However, away from such regions,
and at distances much larger than $\sigv$,
maximizing the \velmod\ likelihood is very similar
to minimizing differences
between TF distances and those inferred from the
velocity field model (the ``Method II'' approach to
velocity analysis; SW, \S~6.4.1).
This suggests that we can find an accurate analytic approximation
to the exact \velmod\ likelihoods for many galaxies. 
Equation~(15) of Paper I is an approximation for
the forward likelihood 
in the simple case when selection effects are neglected
and a constant density field is assumed. 
We have since generalized this result to all
relevant cases and applied it in our calculations,
thereby reducing the run time of the code by a factor of \sm 4. 
The details 
are complex and are given in the Appendix;
we discuss the salient features here.

For each TF galaxy, the velocity
field model yields a ``crossing-point'' distance $w,$
defined implicitly by $w+u(w)=cz,$ where $u(r)$
is the radial component of the predicted peculiar
velocity along the line of sight. 
Similarly, the TF relation defines a distance $d$
implicitly by $5\log d = m-M(\eta)$ (forward) or
$\eta^0(m-5\log d)=\eta$ (inverse). 
Our main result is that 
the forward and inverse single-object
likelihoods are well approximated as normal distributions in
$\ln (d/w),$ but with the
mean value of $\ln d$ offset from $\ln w$ 
by an amount proportional to $\Delta_v^2,$
where $\Delta_v\equiv\sigv/[w(1+u^\prime)],$ and $u'\equiv [du/dr]_{r=w}.$
At large distances, $w\gg \sigv,$ and for $u^\prime\simgt 0,$
$\Delta_v$ is very small and our approximation becomes more accurate. 
In practice we have found that the velocity field is
cold ($\sigv \simeq 130\ \kms;$ \S~\ref{sec:sigmav}),
and $\Delta_v$ is usually small even at distances
as low as 1000 \kms.

We have checked the analytic approximation against the full numerical
integration in a number of regimes.  It fails in regions 
where the velocity
field changes sufficiently rapidly in the vicinity of the crossing point. 
In practice,
this invalidated the approximation in a 30\degs\ cone around
the Virgo cluster, for $w \leq 2700\ \kms.$
More generally, we found that it is
inaccurate for $\Delta_v>0.2$ regardless of the nature of
the velocity field. The
likelihoods for all such objects were always computed
using the full numerical integration.
However, for the remaining \sm 75--80\% of
sample galaxies, we found that
the approximation is remarkably accurate.
The rms difference between the exact and approximate likelihoods 
for such objects is 0.015 in $\like.$
This accuracy is
sufficient to minimize $\like$ at a given $\beta_I$ by
varying the TF parameters and relevant velocity parameters other
than $\beta_I$. 
Once this minimum is found, we re-evaluate $\like$
using the exact numerical probabilities for all
objects. The final  maximum-likelihood
value of $\beta_I$ is derived from these exact values of $\like.$ 
However, this maximum-likelihood value always differed by $<0.01$
from one obtained from the approximations.
Thus, we are confident
that our use of the approximate likelihoods in the parameter
variation procedure has not affected our
maximum likelihood results for $\beta_I.$ 

\section{Selection of the Expanded Sample}
\label{sec:select}
In Paper I we limited our analysis to the local ($\czlg \le 3000\ \kms$) volume.
This constrained us to use
only two TF subsamples of the Mark III
Catalog: the Aaronson \etal\ (1982a; A82) and Mathewson \etal\
(1992; MAT) data sets. The former is a $1.6\mu$m ($H$ band) photometry, 21 cm velocity
width data set; the latter consists of $I$ band CCD magnitudes and a mixture
of 21 cm and optical velocity widths (cf.\ Willick \etal\ 1997a for further
details). The remaining Mark III TF samples contain too few galaxies within
3000 \kms\ to have made their inclusion worthwhile.

Here we increase our redshift limit to $\czlg=7500\ \kms.$ However,
because the
A82 sample itself is badly incomplete beyond 3000 \kms\
(Willick \etal\ 1996), we continue to use the same 300 galaxy, $\czlg \leq 3000\ \kms$ 
A82 subsample used in Paper I. Our MAT TF sample, in contrast, has grown from
538 galaxies in Paper I to 1159 galaxies for the present analysis.
No changes were made in the way we select the MAT galaxies. Specifically,
we continue to apply a (photographic) diameter limit of $1.6$
arcmin and to
require that $\log(a/b)\ge 0.1,$ where $a/b$ is the major to minor axis
ratio. The last requirement excludes objects that are too face-on and
which thus have large velocity width uncertainties.

With the higher redshift limit we now include
the two other 
TF field samples in the Mark III catalog, the Willick (1991; W91PP)
Perseus-Pisces sample and the Courteau-Faber (Courteau 1992, 1996; CF)
Northern sky sample. Both of these data sets consist of $R$ band CCD magnitudes.
W91PP uses the 21 cm velocity widths of Giovanelli \etal\ (1985, 1986)
and Giovanelli \& Haynes (1989), while CF uses optical velocity widths
(Courteau 1997). 
W91PP and CF were originally designed to have uniform
photometric and velocity width properties. 
The mutual consistency of the photometry for these samples
has indeed been verified (Willick 1991; Courteau 1992, 1996). However, their velocity
widths have not been shown to be consistent, and Willick \etal\
(1996, 1997a) found different TF calibrations for the two samples, 
as we will here.

The selection criteria for W91PP and CF are not known as
rigorously as might be hoped.
Both samples are selected to the limit of the UGC catalog---nominally,
therefore, to a photographic diameter limit of 1.0 arcmin. However, the
UGC catalog is known to become increasingly incomplete below about 1.5
arcmin (Hudson \& Lynden-Bell 1991).
This problem was studied
by Willick \etal\ (1996), who found that consistency of W91PP group distance moduli
as measured by the forward and (essentially selection-bias free) inverse
forms of the TF relation
was achieved
with a 
diameter limit of 
1.15 arcmin in evaluating
the selection function. We adopt that result here: we include in the analysis
{\em all\/} W91PP objects down to the UGC limit, but set the formal diameter
limit for evaluating the selection function to 1.15 arcmin. 
As with the
MAT sample, we require $\log(a/b)\geq 0.1.$ The total number of W91PP
galaxies thus included  is 247. 

The selection criteria for the CF sample also included a
photographic magnitude limit of $15.5.$
The expressions derived by Willick (1994) for dealing with this two-limit case
exactly are unfortunately not analytic, and therefore are unsuitable
for \velmod. 
We thus decided to cut the CF sample at a 
larger diameter so that the magnitude limit would
be relatively unimportant, and then use the 
one-catalog selection function corresponding to this larger
diameter.  Specifically, we include
only those CF objects with UGC diameters
$\geq 1.5$ arcmin in the \velmod\ sample, and use a value of 1.6
arcmin in computing the CF selection function to account for
residual incompleteness near the limit. 
As before, we also require 
$\log(a/b)\geq 0.1,$ 
and the total number of CF galaxies
included in the \velmod\ analysis is 170.

\begin{table}[t]
\centerline{\begin{tabular}{l | c l r c }
\multicolumn{5}{c}{{\large TABLE 1}} \\
\multicolumn{5}{c}{Mark III Subsamples Used in the \velmod\ Analysis} \\ \hline\hline
Sample    & Redshift limit & Mag/Diam Limit & $N$ & notes \\ \hline
A82       & $3000\ \kms$ & 14.0 mag & $300$    & a,b \\
MAT       & $7500\ \kms$ & 1.6 arcmin & $1159$ & c \\
W91PP     & $7500\ \kms$ & 1.0 arcmin & $247$  & d \\
CF        & $7500\ \kms$ & 1.5 arcmin & $170$  & e \\ \hline
Total     &        &            & $1876$ &  \\ \hline
\end{tabular}}
\caption{Notes: (a) The A82 sample used in this paper is identical
to that used in Paper I; it is very incomplete beyond 3000 \kms\
and hence was not extended to a higher redshift limit. (b) The
A82 magnitude limit applies to the RC3 catalog blue magnitudes;
see Willick \etal\ (1996) for further details. (c) The
MAT diameter limit applies to the ESO catalog blue photographic
diameters; see Willick \etal\ for further details.
(d) The
formal diameter limit used in calculating the W91PP selection
function was 1.15 arcmin; the diameters in question are
UGC blue photographic diameters. (e) The formal 
diameter limit used
in calculating the CF selection function was 1.6 arcmin;
UGC blue photographic diameters are again used.}
\label{tab:samp}
\end{table}

Our method of assigning diameter limits in the W91PP, CF, and MAT for
computing sample selection functions is far from
satisfactory\footnote{The majority of galaxies in A82 are close enough
that the selection biases are not a serious issue.}.
At distances $\simgt 5000\ \kms$ selection bias becomes an
important effect for the forward relation.  The treatment of selection
effects in \velmod\ 
is in principle rigorous, but is
correct only to the degree that sample selection is properly
characterized. We argued in \S~\ref{sec:inverse-velmod} that we can
test our sensitivity to these problems by carrying out our analysis
using both the forward and inverse relations.  As we shall see, the
results for the two are in excellent agreement, which implies
that our modeling of selection is adequate. 

The total number of galaxies
that enter into the current analysis is 1876. The
TF subsamples, their selection criteria, and the number of objects involved
in each are summarized in Table 1.
As discussed in Paper I, the cluster samples in the Mark III Catalog,
HMCL and W91CL (cf.\ Willick \etal\ 1997a), are not suitable for the
\velmod\ approach, which is tailored to field galaxies, and we do not
include those samples here. We also have elected not to include the
elliptical galaxy portion of the Mark III catalog.

\section{Treatment of the Quadrupole}
\label{sec:quad}
\def\calv{{\cal V}}
In Paper I we presented 
evidence of systematic residuals
from the \iras-predicted velocity field, which could be modeled as
a velocity quadrupole of the form $\bfu_Q=\calv_Q \bfr,$ where
$\calv_Q$ was a traceless, symmetric $3\times 3$ matrix. We argued
that the probable cause of this quadrupole residual was differences
between the true and measured density fields due to shot noise and the
smoothing process, 
at distances
$\simgt 3000\ \kms$. 

We initially treated the 
five independent components of $\calv_Q$ as free parameters 
at each value of $\beta_I$
in our Paper I analysis. 
In our final \velmod\ run, however, we held the quadrupole
fixed at the values of the five components obtained by averaging
their maximum likelihood values at each $\beta_I$, so
that our quadrupole residual would not 
fit out the $\beta_I$-dependent
part of the quadrupole already present in the \iras-predicted velocity
field. The final values of these quadrupole components are given
in Paper I, Table 2, and a
map
of the overall quadrupole flow is presented in Figure 4 
of Paper I. Its rms amplitude averaged over the sky is 3.3\% of 
Hubble flow.
 
The Paper I quadrupole increases linearly
with distance, which is 
the expected signature of a quadrupole
generated at distances beyond the sampled region. 
However, there is no reason to believe that the linear quadrupole extends
beyond $3000\ \kms$. 
A substantial fraction of the Paper I quadrupole
is generated by
mass density determination errors at distances
$3000 \simlt r \simlt 12000\ \kms$
(Paper I, Appendix B).
Such errors will give rise
to a linear quadrupole only at $r\simlt 3000\ \kms.$
In the region coincident with the mass determination errors,
the velocity residual will not have a quadrupole form at all,
and in fact will not be expressible as a divergence-free flow.
Only at distances beyond the region of dominant mass determination
errors will a divergence-free velocity residual reappear,
now with an $r^{-4}$ dependence rather than a linear one (cf.\ Jackson
1976, equation~3.70).

For this paper
we adopt the simplest model consistent with both our
Paper I result and the above considerations. 
We assume that the radial component of the
residual velocity field is given by
\begin{equation}
u_Q(\bfr) = \frac{\calv_Q\,\bfr}{1+\left(r/R_Q\right)^5}\cdot\hat\bfr\,, 
\label{eq:newquad}
\end{equation}
where $\hat\bfr\equiv\bfr/r.$ 
In equation~(\ref{eq:newquad}), $\calv_Q$ is the {\em same\/} traceless,
symmetric $3\times 3$ matrix as was derived in Paper I. However,
we have introduced a new quantity, $R_Q,$ that parameterizes
the cutoff scale of the linear quadrupole. For $r\ll R_Q,$
we recover the Paper I quadrupole exactly. For $r\gg R_Q,$
we obtain the $r^{-4}$ quadrupole expected at large distances.
The transition
between them is smooth but 
rapid, so there is only a small
region, with $r\approx R_Q,$ in which $u_Q(\bfr)$ does not behave
like a quadrupole. This is a desirable feature, for it minimizes
the volume in which our residual velocity field has divergence.
We will determine the value of $R_Q$ through maximum likelihood in \S~5.3,
and justify our modeling of the quadrupole \aposteriori\ in
\S~\ref{sec:residuals}, where we show that the \iras\ velocity
field, with the quadrupole
included, gives an acceptable fit to the TF data.

\section{Results}
\label{sec:results}
In this section we present the main results of applying \velmod\ to
the 1876-galaxy, $\czlg\leq 7500\ \kms$ subsample described
in \S~\ref{sec:select}.  In \S~\ref{sec:TFscatter}, we search for a luminosity
dependence to the TF scatter. We show the results for $\beta_I$
without allowing for an external quadrupole
in \S~\ref{sec:noquad}.
In \S~\ref{sec:quad-RQ} we find the value of $R_Q$ to use in the
quadrupole formula, equation~\ref{eq:newquad}.  We
then use this quadrupole in an analysis of the small-scale velocity
dispersion in \S~\ref{sec:sigmav}, where we give our final results for
$\beta_I$.  The robustness of this result to subsample is discussed in
\S~\ref{sec:robust}, and to smoothing scale in
\S~\ref{sec:500-smooth}. 

\subsection{Establishing the luminosity/velocity-width dependence of the TF scatter}
\label{sec:TFscatter}
Giovanelli \etal\ (1997) and Willick \etal\ (1997a) have pointed out
that in some samples, the TF scatter is a function of luminosity.  
In Paper I, we modeled the velocity-width dependence of the
forward TF scatter 
as having the form $\sigtf(\eta)=\sigma_0 -g_f\eta$, and adopted the 
Willick \etal\ (1997a) values of 
$g_f=0.14$ for A82 and $g_f=0.33$ for MAT.
Here we similarly model the luminosity-dependence
of the inverse TF scatter by $\sigma_\eta(M) = \sigma_{\eta,0} + g_i(M -
\overline M)$, where $\overline M$ is 
the mean absolute magnitude for the sample\footnote{The absolute magnitudes
were calculated for this purpose as $M=m-5\log\czlg$, so that they are
independent of $\beta_I$.}, but determined the $g_f$ and $g_i$ through
maximum likelihood, as follows.

First, we ran a preliminary set
of \velmod\ runs, both forward and inverse, in which the $g_f$
and the $g_i$ were treated as free parameters at each value
of $\beta_I.$ 
These runs demonstrated that there was negligible cross-talk between
the $g_f$ ($g_i$) and any other parameter of interest, in particular,
$\beta_I$. 
We thus used these preliminary
runs to establish their values, and then held them
fixed for all subsequent \velmod\ runs. The preliminary runs
employed the simplest
velocity models: no quadrupole, $\bfwlg$ fixed at its Paper I,
no quadrupole value, and $\sigma_v$ fixed at $150\ \kms$ without
allowance for a density dependence (cf.\ \S~\ref{sec:sigmav}).

We imposed an additional constraint on the $g_f$ and $g_i.$
The TF relation implies that
\begin{equation}
g_i = \frac{d\sigeta}{dM} \approx - \frac{d(\sigtf/b)}{d(b\eta)}=
- \frac{1}{b^2}\frac{d\sigtf}{d\eta} = \frac{g_f}{b^{2}}\,,
\label{eq:gigf}
\end{equation}
where $b$ is the forward TF slope for the sample in question.
We can regard
$g_f$ and $g_i$ as well-determined from the data to the degree that
this relation holds. We found
that the A82 and MAT samples satisfied equation~(\ref{eq:gigf}),
and thus adopted the values of $g_f$ and $g_i$ determined
from the preliminary \velmod\ runs for those samples. For W91PP, however,
$g_f$ and $g_i$ obtained from those runs had opposite signs and were thus
inconsistent with equation~(\ref{eq:gigf}).  We interpret this to mean
that there is no significant luminosity or velocity width dependence
of the W91PP TF scatter, 
a conclusion also reached
by Willick \etal\ (1997a). We thus set $g_f=g_i=0$ for W91PP.

For CF, the preliminary $g_f$ and $g_i$ were both positive, but $g_i$ was
significantly smaller than $b^{-2} g_f$.
CF uses optical widths, as does MAT. 
MAT shows the
strongest signal of a luminosity-dependent scatter, which we conjecture
is a consequence of optically-measured widths. We thus
assigned CF the same value of $g_f$ as was found
by maximum likelihood for MAT. 
For the CF $g_i,$ we took the mean of its maximum likelihood value 
and the value inferred from equation~(\ref{eq:gigf}) given
the adopted value of $g_f.$

\begin{table}[t]
\centerline{\begin{tabular}{l | r r }
\multicolumn{3}{c}{{\large TABLE 2}} \\
\multicolumn{3}{c}{Luminosity/Velocity-width} \\
\multicolumn{3}{c}{Dependence of TF Scatter} \\ \hline\hline
Sample    & \multicolumn{1}{c}{$\;g_f$}&\multicolumn{1}{c}{$\;g_i$}\\ \hline
A82       & $-0.24$ & $-0.0021$ \\ 
MAT       & $0.35$  & $0.0055$  \\
W91PP$^a$ & $0.00$  & $0.0000$  \\
CF        & $0.35$  & $0.0030$  \\ \hline
\end{tabular}}
\caption{Notes: (a) No significant luminosity/width dependence of scatter was
detected.}
\label{tab:gigf}
\end{table}
We summarize the results of this exercise in Table~\ref{tab:gigf}. Column (1)
gives the sample name, while columns (2) and (3) list the adopted
values of $g_f$ and $g_i$ respectively. Note that the MAT value of $g_f$
is very close to that found by Willick \etal\ (1997a). This is
not surprising, as MAT is the only sample for which the luminosity
dependence of scatter has a strong, unambiguous signal. The A82 value
of $g_f,$ however, not only differs from the Willick \etal\ (1997a)
value, but is negative, signifying a scatter {\em increase\/} with
increasing luminosity. The physical reason for this is unclear, but
the consistency between $g_f$ and $g_i$ for A82 suggests that the
effect is real. 
However, these choices have no meaningful effect
on the derived values of $\beta_I$ (as mentioned above) or any other important
quantity discussed in this paper. 
Values of $\sigtf$
or $\sigeta$ quoted later in this paper refer to their values at
$\eta=0$ and $M=\overline{M}$ respectively. 

\subsection{No-quadrupole results}
\label{sec:noquad}
After adopting the values of $g_f$ and $g_i$ in Table 2, we reran
forward and inverse \velmod\ with no quadrupole and $\sigma_v$
fixed at 150\ \kms. For these runs, we also fixed the LG random
velocity vector $\bfwlg$ at the value determined in Paper I for the
no-quadrupole case. 

\begin{figure}[t!]
\plottwo{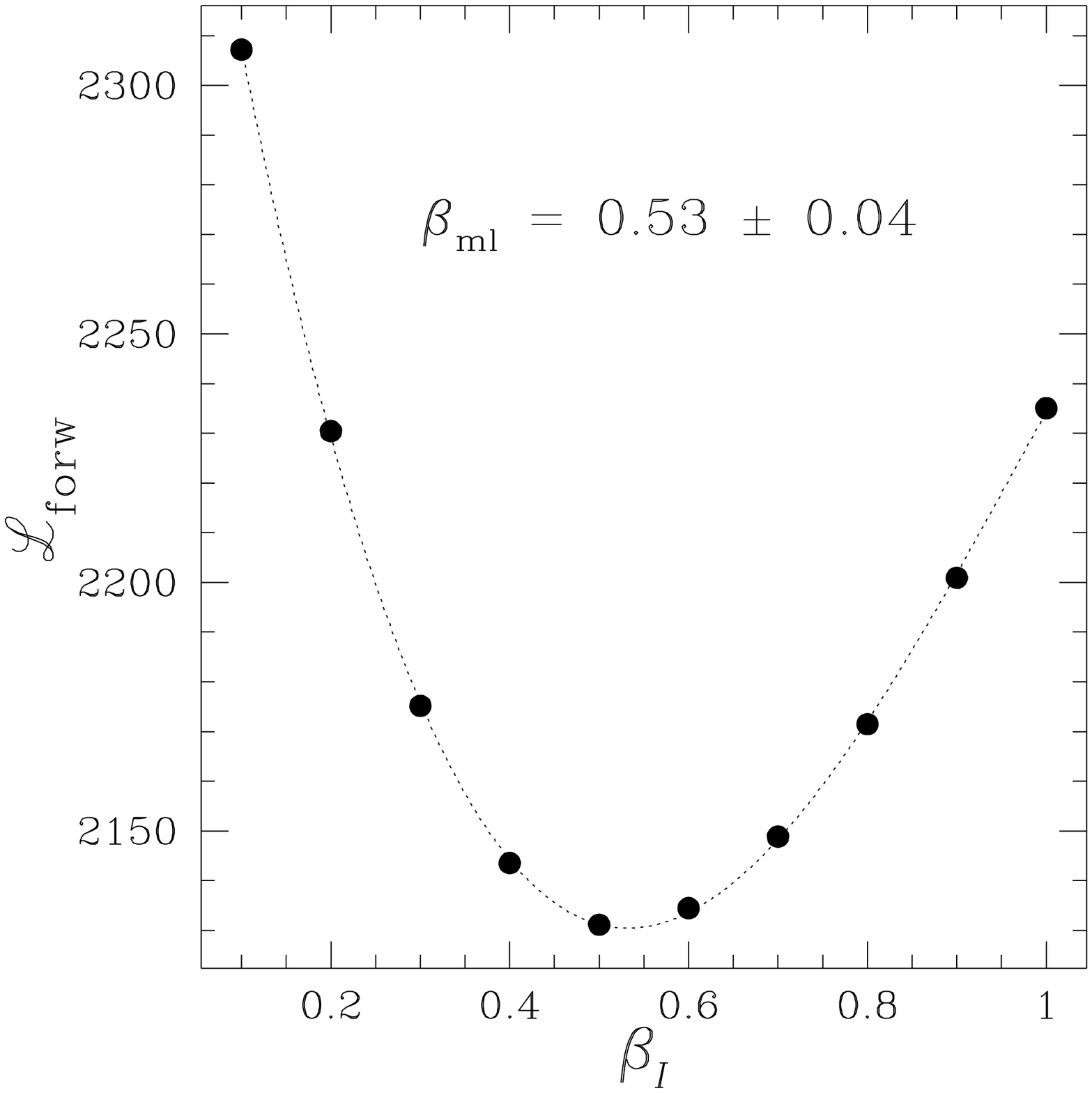}{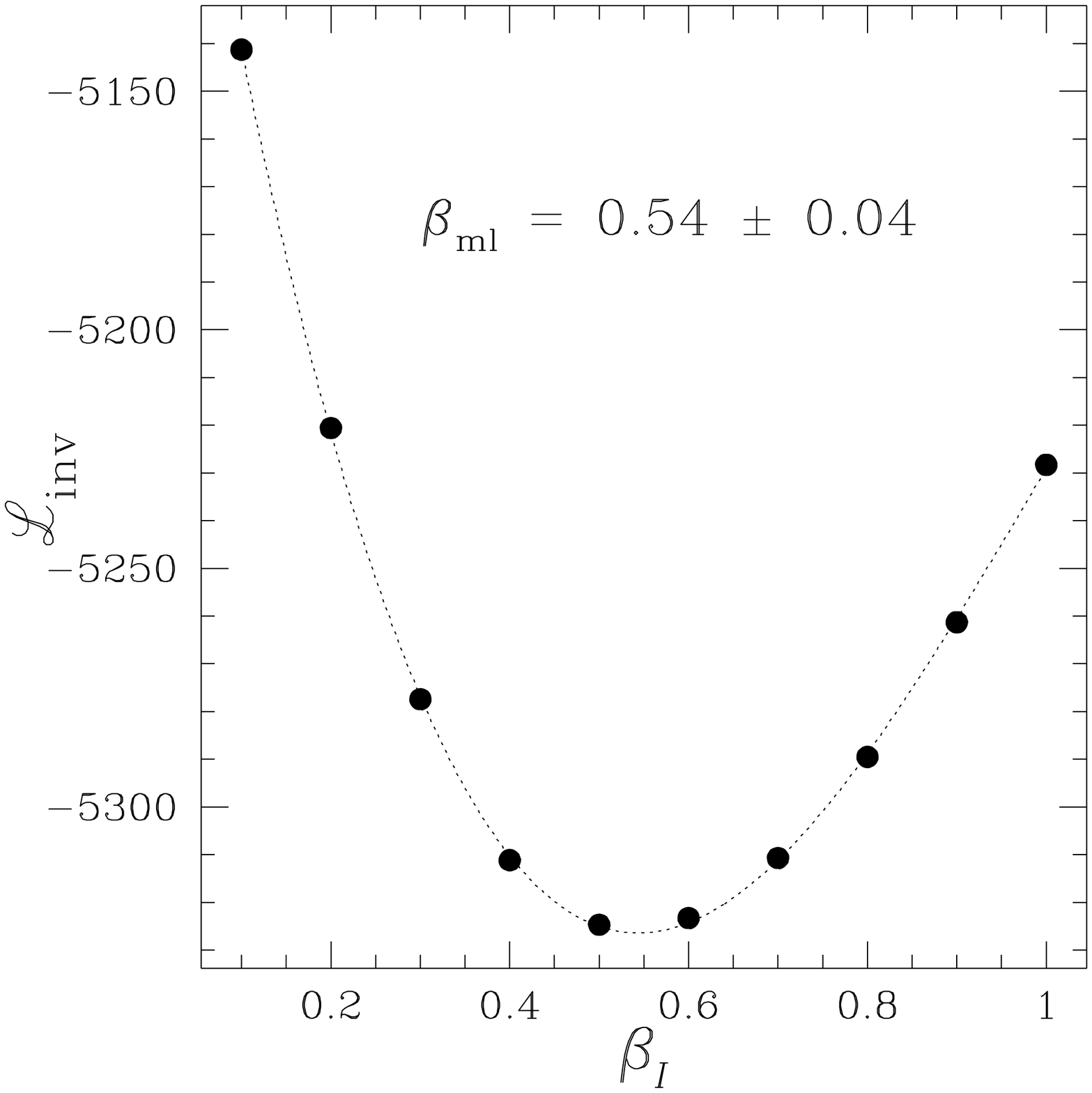}
\caption{{\small The \velmod\ likelihood statistic $\like$ 
as a function of $\beta_I$ for the no-quadrupole case.
Results are shown for both the forward (left panel)
and inverse (right panel) TF relations. 
In both cases the velocity noise $\sigv$ has
been held fixed at 150 \kms\ and the LG random velocity
vector $\bfwlg$ has been fixed at its Paper I, no-quadrupole value. 
The dotted curves are the cubic fits to the $\like(\beta_I)$ points.
The maximum likelihood values of $\beta_I$ are indicated
on the plots as $\beta_{{\rm ml}}.$}}
\label{fig:4snq}
\end{figure}
The results 
are shown in Figure~\ref{fig:4snq} for the forward (left panel) and inverse (right panel)
TF relations\footnote{The absolute values of the forward and inverse likelihood
statistics are quite different because the former derives
from a probability density in apparent magnitude, the latter
from a probability density in the width parameter $\eta.$}. 
The full likelihood versus $\beta_I$ curves are quite 
similar for the forward and inverse TF relations.  In particular, the
maximum likelihood values 
of $\beta_I$ differ by only $0.01,$ which is insignificant
given that the \onesigma\ error in $\beta_I$ is $0.04.$ 
As we shall see, forward
and inverse give essentially identical results for $\beta_I$ for all
\velmod\ runs. 
Agreement between
the forward and inverse results means that 
our approximate treatment of the selection
and luminosity functions have no meaningful effect on
$\beta_I$ (see the discussion in \S~\ref{sec:inverse-velmod}).

Finally, note that the value of $\beta_I$ obtained here for the
no-quadrupole case is very close to the value of $\beta_I=0.56$ obtained
in Paper I for the no-quadrupole case. Thus, more than doubling
the number of sample objects and extending the redshift limit
from $3000$ to $7500\ \kms$ has had essentially no effect on
$\beta_I$ (other than shrinking the error bar). 
% for the no-quadrupole case. 
We will see below that the same
is true when the quadrupole is included.

\subsection{Determining the quadrupole cutoff scale}
\label{sec:quad-RQ}
As described in \S~\ref{sec:quad}, we adopted a quadrupole velocity residual,
equation~(\ref{eq:newquad}), 
that agrees with the Paper I quadrupole at small distances, but
changes smoothly from a linear to an $r^{-4}$ quadrupole for $r>R_Q.$
To determine the value of $R_Q$ we carried out a series of \velmod\
runs, both forward and inverse, with values of $R_Q$ ranging from
$100\ \kms$ (which essentially means no quadrupole) 
to $15,000\ \kms$ (which amounts to the Paper 1 quadrupole throughout
the sample volume).  In each of these runs, $R_Q$ was held fixed, as were
$\sigv$ at $150\ \kms$ and $\bfwlg$ at its best-fit Paper I value
when the quadrupole was included. Only $\beta_I$ and the
12 TF parameters were varied in each run of a given $R_Q.$

\begin{figure}[t!]
\centerline{\epsfxsize=4.5 in \epsfbox{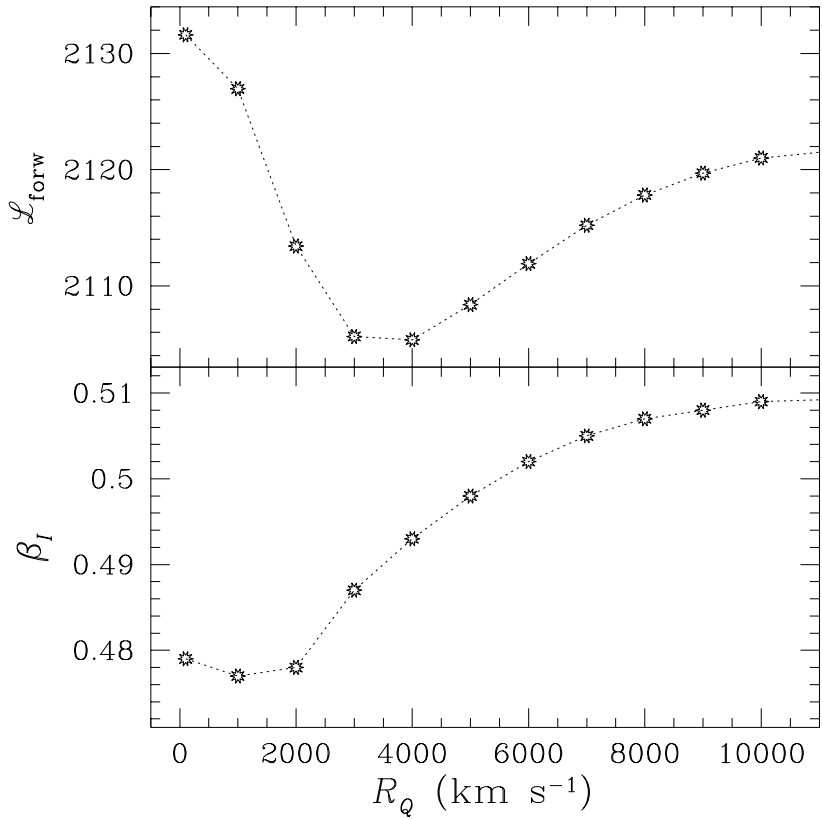}}
\caption{{\small The best forward TF likelihood $\likeforw$ (upper panel), 
and the corresponding
maximum likelihood value of $\beta_I$ (lower panel) plotted as a function of the
quadrupole scale $R_Q$ (cf.\ equation~[\ref{eq:newquad}]). 
For these runs the velocity noise was held fixed at $\sigv=150\ \kms$
and the LG random velocity vector $\bfwlg$ at its Paper I, quadrupole value.
Note the strong likelihood maximum for $R_Q=3000$--4000 \kms. The
corresponding value of $\beta_I$ is 0.49.}} 
\label{fig:rq_forw}
\end{figure}

\begin{figure}[t!]
\centerline{\epsfxsize=4.5 in \epsfbox{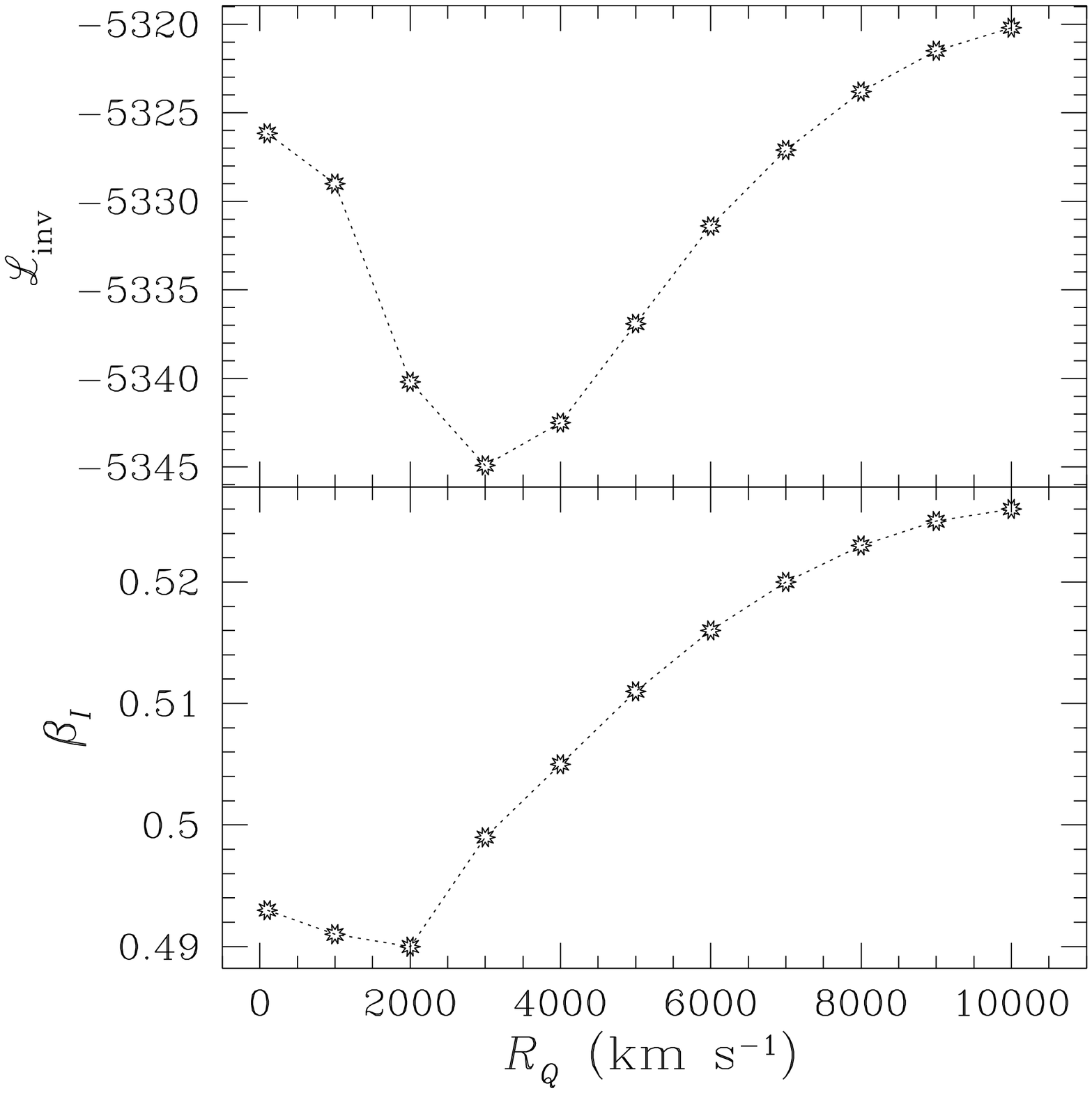}}
\caption{{\small As in the previous figure, for the inverse TF relation.}}
\label{fig:rq_inv}
\end{figure}

Figures~\ref{fig:rq_forw} and~\ref{fig:rq_inv} show the results 
of these runs for the forward
and inverse relations respectively. In each figure, the upper
panel shows $\like$ 
%the maximum likelihood value of the
%likelihood statistic $\like$ (recall that the lower $\like,$
%the higher the likelihood)
versus $R_Q$, while the lower
panel shows the maximum likelihood value of $\beta_I$
versus $R_Q.$ We show the results only out to $R_Q=10,000\ \kms,$ as
for larger $R_Q$ neither $\beta_I$ nor $\like$ changed appreciably.

%Two important points emerge
%from these figures. First, although $\beta_I$ varies systematically
%with $R_Q,$ the variation is small. 
Note that $\beta_I$ is very insensitive to $R_Q$; over the entire range of
$R_Q$ considered, $\beta_I$ changes only by \sm 0.03, or
less than \onesigma.  There is a well-defined likelihood
maximum (minimum of $\like$) 
%for both the forward and inverse relations. The likelihood maximum is
at $R_Q = 3000\ \kms$ for
the inverse case and at $R_Q = 4000\ \kms$ for the forward case.
Note that $R_Q\leq 2000\ \kms$ and $R_Q\geq 5000\ \kms$
are strongly disfavored in both cases, while $R_Q = 3500 \ \kms$ is
consistent with both, so we adopt the latter value for the remainder
of the paper.  The
maximum likelihood values of $\beta_I$ are very close to
$0.50$ for this value of $R_Q$, for both
forward and inverse VELMOD. 

The value of $R_Q = 3500$ \kms\ signifies that the Paper I 
quadrupole cuts off strongly beyond this distance.   We will see in
\S~\ref{sec:residuals} that the resulting velocity field is an adequate
fit to the data.  We can therefore conclude that 
the \iras\ versus true mass differences
arising from the smoothing/filtering procedure that 
dominate the velocity prediction errors are concentrated
in the range \sm 3000-5000 \kms. 
%Such mass differences
%give rise
%to a linear quadrupole at small ($\simlt 3000\ \kms$) distances
%and an $r^{-4}$ quadrupole at $r \simgt 5000\ \kms.$
%{\bf Again, we can say this only if we convince ourselves a posteriori
%that our velocity field model is a good fit.} 

At $R_Q = 100\ \kms$, corresponding to essentially no quadrupole, we
find $\beta_I = 0.48$ for the forward relation, which differs from the
no-quadrupole value of 0.53 found in \S~\ref{sec:noquad}.
These differ because different values of $\bfwlg$ were used:
In \S~\ref{sec:noquad} 
we fixed $\bfwlg$ at the Paper I, no-quadrupole
best-fit value, while here, we used 
the Paper I
quadrupole best-fit value. 
%We could carry out a more general exercise
%in which both $\bfwlg$ and $R_Q$ vary. However, because 
The values of $\like$ in the two cases are similar, however, implying
that we have limited sensitivity to $\bfwlg$ in the likelihood
analysis. There is moderate covariance between $\bfwlg$
and $\beta_I$ when a quadrupole is not used to describe the local
flow field. With the quadrupole added, however, this covariance is
much reduced. Stated another way, when the quadrupole is modeled,
$\bfwlg$ is both smaller in amplitude and better determined. 

\subsection{The small-scale velocity dispersion}
\label{sec:sigmav}
\begin{figure}[t!]
\centerline{\epsfxsize=4.5 in \epsfbox{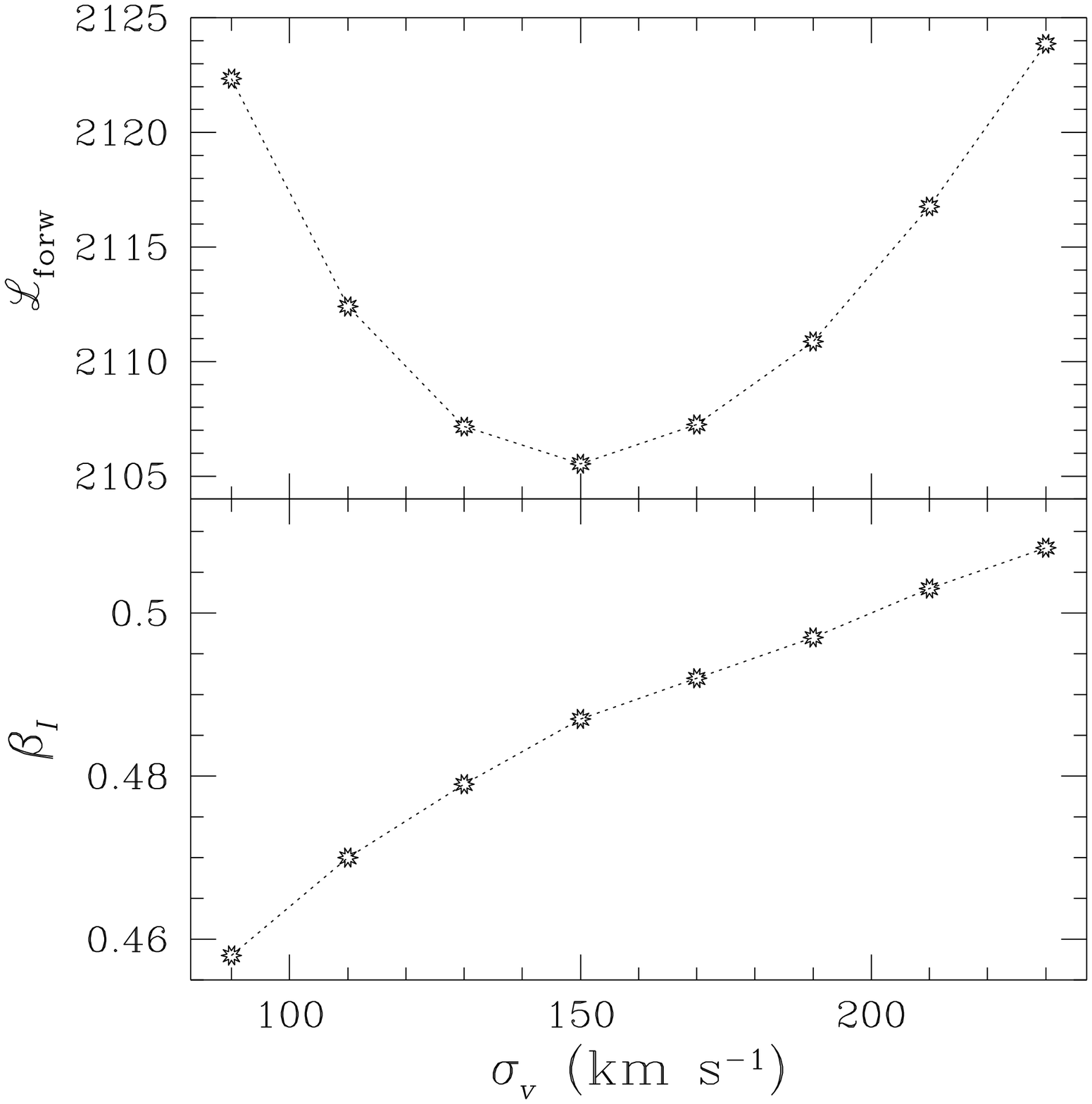}}
\caption{{\small The best forward TF likelihood $\likeforw$ (upper panel), 
and the corresponding
maximum likelihood value of $\beta_I$ (lower panel) plotted as a function of the
small-scale velocity dispersion $\sigma_v.$
For these runs the quadrupole scale was fixed at $R_Q=3500\ \kms,$
and the LG random velocity vector at its Paper I, quadrupole value.
The likelihood statistic yields
$\sigv = 150 \pm 20\ \kms$. 
Within this favored range $\beta_I$
varies trivially, from 0.48--0.49.}}
\label{fig:sigv_forw}
\end{figure}

\begin{figure}[t!]
\centerline{\epsfxsize=4.5 in \epsfbox{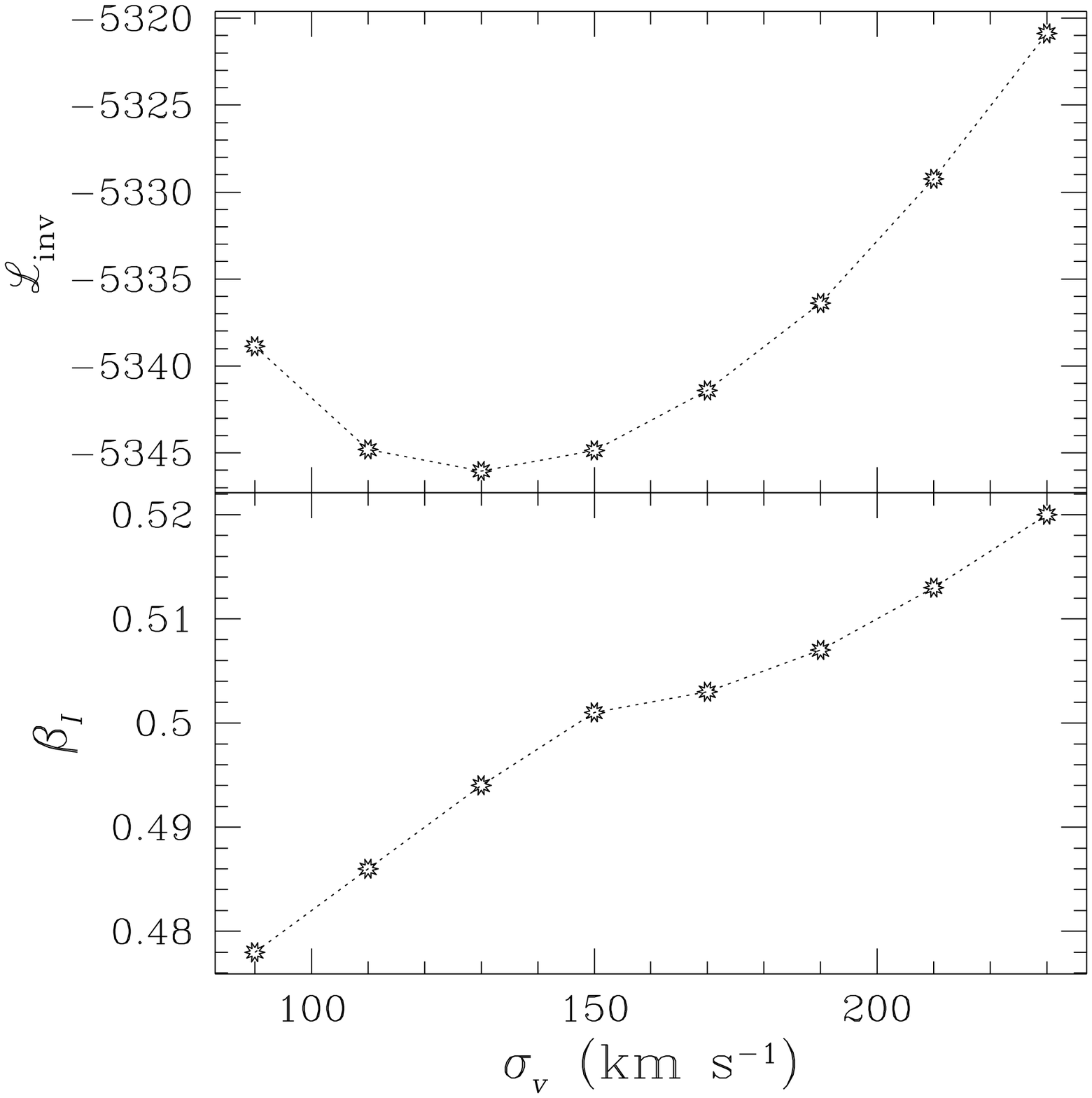}}
\caption{{\small As in the previous figure, for the inverse TF
relation.}}
\label{fig:sigv_inv}
\end{figure}

In the \velmod\ runs described up to now, the small-scale velocity
dispersion $\sigma_v$ has been fixed at $150\ \kms$,
% This was 
%adopted merely as 
a useful round number with which to establish 
the values of %unrelated 
%quantities such as the 
$g_f$, $g_i,$ and $R_Q.$
Having done so, we ran a series of \velmod\ runs
for a range of fixed values of $\sigma_v,$ with $R_Q$
fixed at $3500\ \kms$ and $\bfwlg$ fixed at its
Paper I, quadrupole value. In each run, $\beta_I$ and
the 12 TF parameters were varied to maximize likelihood.
Figures~\ref{fig:sigv_forw} and
\ref{fig:sigv_inv} show the results
for the forward and inverse relations respectively. In each case,
we plot the maximum likelihood values of $\beta_I$ and the 
corresponding $\like(\beta_I)$ versus $\sigma_v.$ 

%[MS: I changed your language of ranges for $\sigv$ to 1 sigma error bars.]
There is a weak systematic variation of $\beta_I$ with $\sigma_v,$
amounting to less than $0.05$
over the full range of $\sigma_v$ considered.  The likelihood reaches
a clear maximum at $\sigma_v = 150 \pm 20\ \kms$ for the forward
relation, and $130 \pm 20\ \kms$ for the inverse relation. 
%More importantly,
%there is again a range of $\sigma_v$ for which the likelihood
%is highest, corresponding to \sm 130--170 \kms\ for the
%forward relation and to 110--150 \kms\ for the inverse relation.
%Values of $\sigv$ less than \sm 100 \kms\ or greater than \sm 180 \kms\
%are strongly ruled out by these results.
Over the \onesigma\ errorbar,
%Within the favored ranges of $\sigv,$, 
the maximum likelihood
values of $\beta_I$ only vary by 0.01, much less than the statistical
error on this quantity. 
%trivially, from 0.48--0.49 (forward)
%and 0.49--0.50 (inverse). Thus, the systematic trend of $\beta_I$
%with $\sigma_v$ is statistically unimportant (less than \onesigma).
Thus 
%Stated another way, 
there is little covariance between $\beta_I$ and $\sigv$.  
%within the allowed range of $\sigma_v.$ 

The value of $\sigma_v$
found here is consistent with the maximum likelihood value
$\sigma_v=125 \pm 20\ \kms$ found in Paper I. This is
reassuring, though not surprising; as discussed in Paper I,
$\sigma_v$ is primarily determined at small distances, $< 3000\
\kms$, where
its effect on the overall variance is comparable to that of the
TF scatter. 
%Beyond about 3000 \kms, $\sigma_v$ has a negligible
%effect on the variance and thus cannot be constrained by
%data points at these large distances. 
%The preferred value of $\sigma_v$
%is larger here than it was in Paper I, at least for the forward
%case. In the next section we discuss a possible reason for this.

%[MS: I've taken out this section heading, as this next discussion is
%really a continuation of what the previous section.] 
%[JW - I think it deserves at least a subsubsection.]
\subsubsection{Density-dependence of $\sigma_v$}
\label{sec:sigmav-delta}
Strauss, Ostriker, \& Cen (1998) and Kepner, Summers, \& Strauss
(1997) showed that the small-scale velocity dispersion is an
increasing function of local density.  
%We mentioned in Paper I that $\sigv$ most
%likely varies with galaxy density
In Paper I, we chose to neglect such variation, 
the only exception being our ``collapsing'' of 20 Virgo
cluster galaxies by assigning them 
%radial velocities 
redshifts 
equal
to the cluster mean (cf.\ Paper I, \S 4.3).
For this paper %, with our larger sample and
%faster code, 
we attempt to detect a density-dependence
of $\sigma_v$ through the likelihood analysis. 
%Motivated
%by the recent work of Strauss, Ostriker, \& Cen (1997),
We adopt a model of the form
\begin{equation}
\sigma_v(\delta_g) = \sigma_{v,1} + f_\delta\,(\delta_g-1)\,,
\label{eq:sigvdelta}
\end{equation}
where $\delta_g$ is at the same smoothing as was assumed for the
\iras\ velocity field calculation. 
We take $\delta_g=1,$
rather than $\delta_g=0,$ as the zero point for our model 
because most TF sample objects lie in relatively high-density 
environments (the mean value of $\delta_g$ for the full TF
sample is $\sim 0.8$). 
%There is thus less
%covariance between $f_\delta$ and the zero point than
%there would have been had we adopted $\delta_g=0$ as a zero point.
%[MS: This seemed too minor a point to be worth these sentences.]
%[JW: I disagree; saying nothing leaves the normalization hanging
%and the reader will question it; I do agree that one rather than
%two sentences suffices.]

\begin{figure}[t!]
\centerline{\epsfxsize=4.5 in \epsfbox{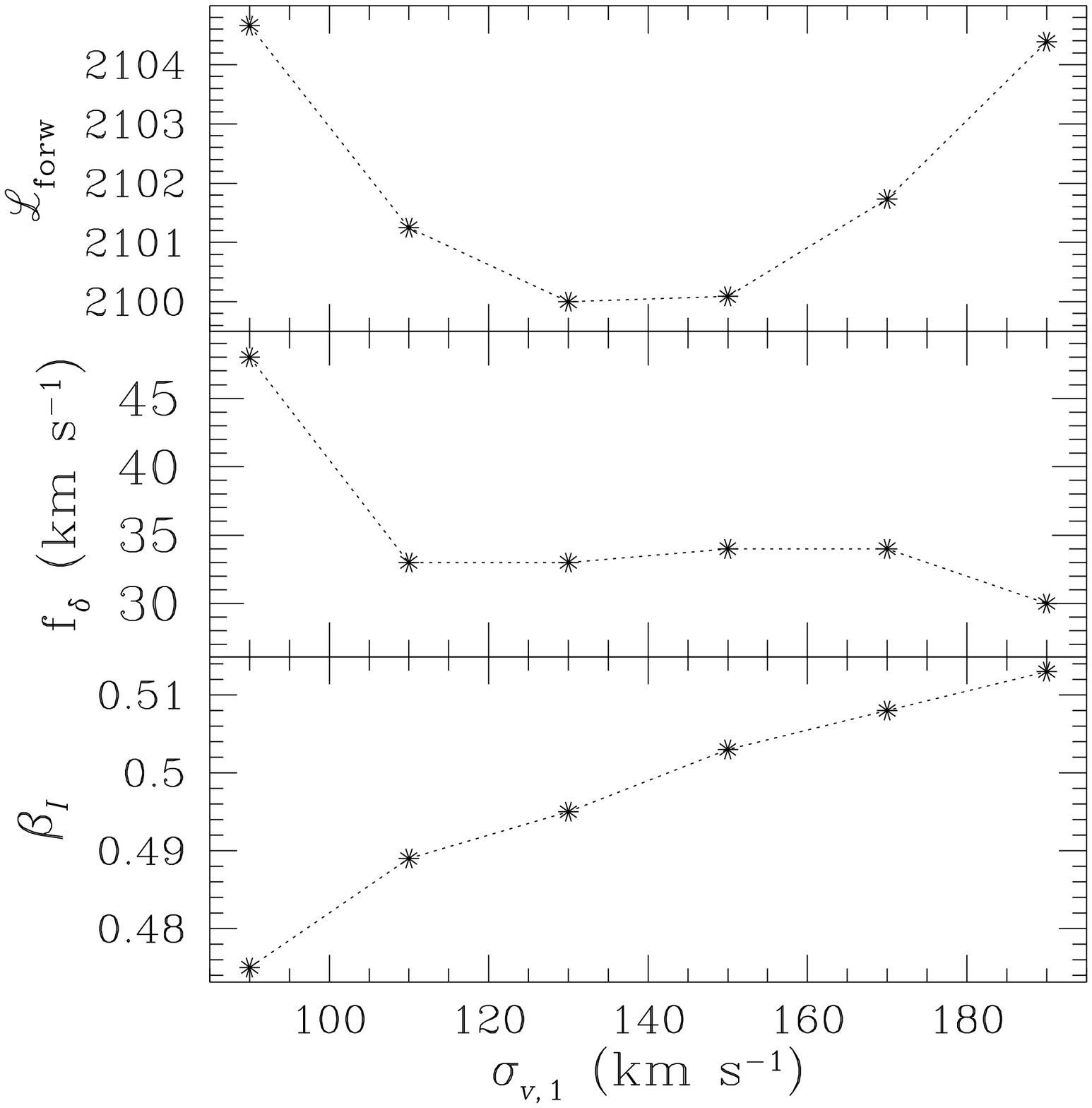}}
\caption{{\small The results of a series of forward
\velmod\ runs in which the small-scale velocity dispersion
is treated as a linear function of galaxy density
(cf.\ equation~[\ref{eq:sigvdelta}]). The bottom panel shows
the maximum likelihood value of $\beta_I,$ the middle
panel the corresponding value of $f_\delta,$ and 
the upper panel the corresponding value of the
likelihood statistic as a function of $\sigma_{v,1}.$
The Paper I quadrupole with $R_Q=3500\ \kms$ was used.}}
\label{fig:bflike_forw}
\end{figure}
We carried out a series of forward and inverse \velmod\ runs for a range of
values of $\sigma_{v,1}.$ 
In each run $\beta_I,$ the
twelve TF parameters, and $f_\delta$ were treated as free parameters.
We continued to hold $R_Q$ fixed at $3500\ \kms.$ The results of this
exercise are shown in Figure~\ref{fig:bflike_forw} for the forward
relation, which plots 
%(The inverse results look substantially the same except as
%described below.)
the maximum likelihood
values of $\beta_I,$ and the corresponding values of
$f_\delta$ and $\likeforw$
%are plotted %in the lower, middle, and upper
%panels of each plot 
as a function of $\sigma_{v,1}.$ 
%The plot shows that 
Allowing for a density-dependent
velocity dispersion has no significant effect on
our derived $\beta_I,$ which remains
very close to $0.50$ near the minimum of $\likeforw.$

The best likelihood is achieved for
$\sigma_{v,1}\simeq 140\ \kms,$ very similar
to the value of the 
invariant $\sigma_v$ for which likelihood was
maximized (compare with Figure~\ref{fig:sigv_forw}). For $\sigma_{v,1}$ in the
range favored by the likelihood statistic, $140\pm 25\ \kms$, 
$f_\delta$ is remarkably constant at $33$--$34$ \kms.
The minimum value of $\likeforw$ in Figure~\ref{fig:bflike_forw}
is $5.5$ points smaller than its minimum value for an
invariant $\sigma_v,$ corresponding to an increased likelihood
of the fit by a factor of $e^{(5.5-1)/2}\approx 9.5,$ 
a $2.1\,\sigma$ result.
%to phrase this is that this is a $\sqrt{4.5} \approx 2$ sigma result.}
For the inverse
relation (not shown) $f_\delta = 36\ \kms$ when the best likelihood
is achieved for $\sigma_{v,1} = 120\pm 25\ \kms,$ 
and the best likelihood is greater than for the invariant $\sigv$
case by a factor of \sm 25, a $2.5\,\sigma$ result.  
We have thus 
detected a significant variation of 
velocity dispersion with density. 
%The inverse results differ
%from the forward in that likelihood is maximized for
%$\sigma_{v,1} = 110\ \kms,$ somewhat smaller than the
%forward value. However, $\sigma_{v,1}=130\ \kms$ is
%within \onesigma\ of maximum likelihood in
%both cases.   
%[MS: Again, let's quote real error bars.] 
To a good approximation we may summarize
these results (now normalizing to $\delta_g = 0$) as 
$\sigma_v = [100 \pm 25 + 35\,\delta_g]\ \kms.$

The value of $\sigma_v$ for galaxies in a mean density environment
is very small, consistent with the conclusions of Paper I, Davis,
Miller, \& White (1997), Strauss \etal\ (1998), and papers referenced therein. 
%This strengthens our conclusion from Paper I,
%in which we found $\sigv=125\pm 20\ \kms,$
%that the velocity dispersion for field galaxies is very small.
%[MS: We already said this!]
%It should be recalled that 
The quantity $\sigma_v$ is the quadrature sum
of \iras\ error and true velocity noise (cf.\ Paper I, \S~3.2).  We
estimated the former to be 
$\sim 70$ \kms\ in Paper I, so the
true 1-D velocity noise is only about 50--70
\kms\ in mean-density environments. The flow field of galaxies
is remarkably cold. 

\begin{figure}[t!]
\plottwo{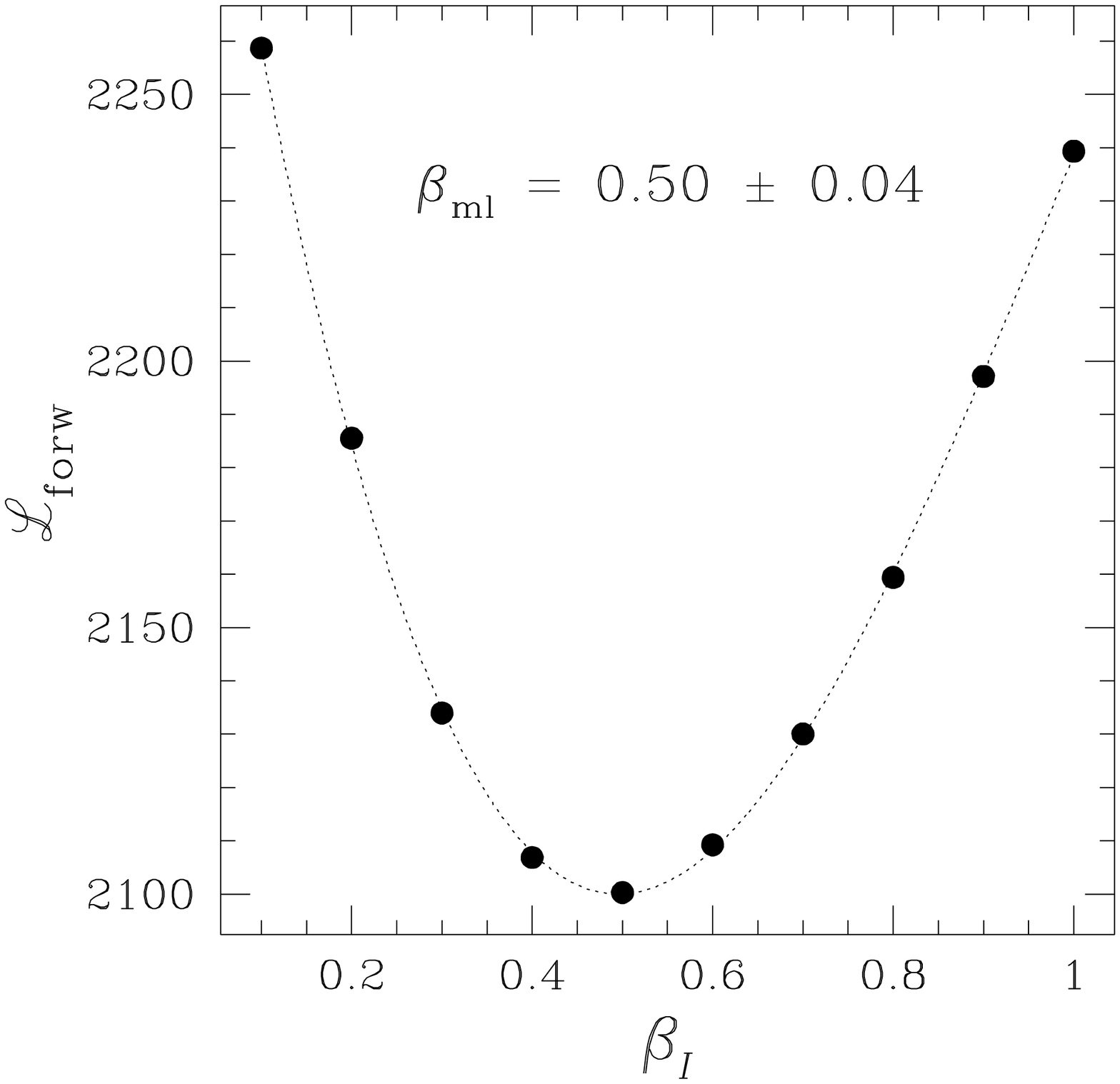}{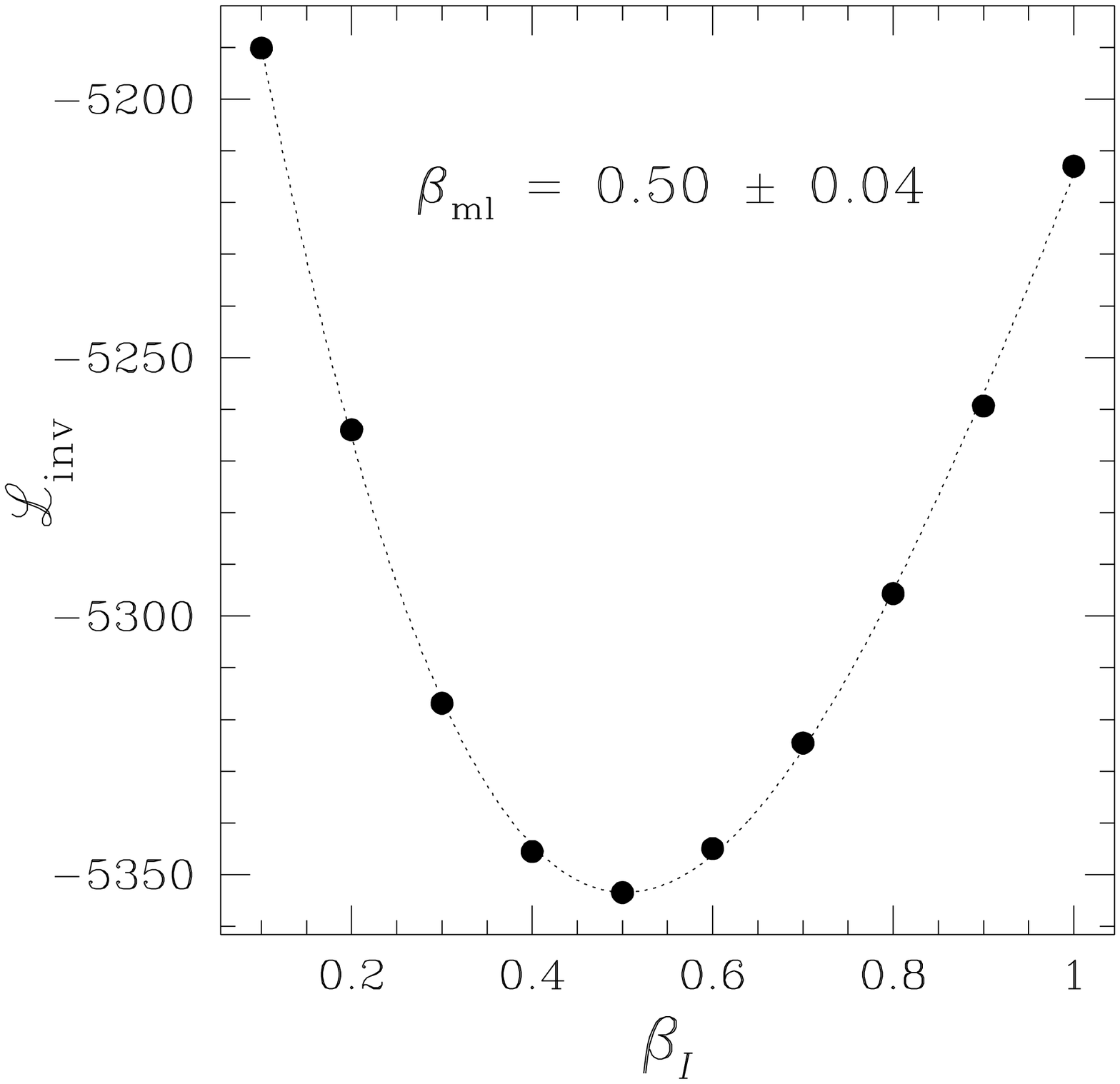}
\caption{{\small The \velmod\ likelihood statistic $\like$
as a function of $\beta_I$ for the forward TF (left panel)
and inverse TF (right panel) relations. 
The maximum likelihood values of $\beta_I$ are indicated on
the plots. For these \velmod\ runs, the 300 \kms-smoothed \iras\
velocity field, 
the Paper I quadrupole with $R_Q=3500\ \kms,$
and the LG random velocity
vector $\bfwlg$ fixed at its Paper I value were used. A density-dependent
velocity dispersion (cf.\ equation~[\ref{eq:sigvdelta}]) was used in these
runs, with $f_\delta$ treated as a free parameter
at each $\beta_I.$ A value of $f_\delta\approx 35\ \kms$ was
obtained for $\beta_I=0.5$ in both cases. For the forward TF run the results
for $\sigma_{v,1}=130\ \kms$ are plotted, while for the inverse
TF run the results for $\sigma_{v,1}=110\ \kms$ are shown.}} 
\label{fig:ddsfi}
\end{figure}
In Figure~\ref{fig:ddsfi} we plot the forward and  inverse likelihood 
statistics versus $\beta_I.$ The plots are done for 
$\sigma_{v,1}=130\ \kms$ in the forward
case, and $\sigma_{v,1}=110\ \kms$ for the inverse case. 
%The plots would
%look nearly the same if we chose $\sigma_{v,1}=150\ \kms$ for the forward
%case and $\sigma_{v,1}=130\ \kms$ for the inverse case. 
As in Figure~\ref{fig:4snq}, 
%can be seen,
%aside from the different likelihood scale, 
the forward
and inverse curves are
almost identical, and the resultant maximum likelihood values of
$\beta_I$ are the same to within $0.01$ (see 
Table~\ref{tab:breakdown}).
This tells us that any errors we may
have made in modeling sample selection and luminosity functions
have had little or no effect on the quantities of interest.

\subsection{Breakdown by sample and redshift}
\label{sec:robust}
\begin{table}[t]
\centerline{\begin{tabular}{l | l l r }
\multicolumn{4}{c}{TABLE 3: Breakdown by Sample and Redshift} \\ \hline\hline
%\multicolumn{4}{c}{{\large Breakdown by Sample and Redshift
%$^{a,b}$
Subsample    &\multicolumn{1}{c}{$\beta_I$ (forward)}&
\multicolumn{1}{c}{$\beta_I$ (inverse)} &\multicolumn{1}{c}{$N$} \\ \hline
A82       & $0.477 \pm 0.062$ & $0.486 \pm 0.061$ & 300 \\
MAT       & $0.518 \pm 0.052$ & $0.533 \pm 0.052$ & 1159 \\
W91PP     & $0.411 \pm 0.159$ & $0.386 \pm 0.148$ & 247 \\
CF        & $0.488 \pm 0.107$ & $0.478 \pm 0.117$ & 170 \\ \hline
$cz\leq 1500\ \kms$&$0.515 \pm 0.059$&$0.510 \pm 0.056$& 327 \\
$1500 < cz\leq 3000\ \kms$&$0.542 \pm 0.065$&$0.532 \pm 0.066$& 564 \\
$3000 < cz\leq 4500\ \kms$&$0.428 \pm 0.084$&$0.473 \pm 0.088$& 370 \\
$4500 < cz\leq 6000\ \kms$&$0.376 \pm 0.096$&$0.381 \pm 0.094$& 422 \\
$6000 < cz\leq 7500\ \kms$&$0.594 \pm 0.173$&$0.734 \pm 0.197$& 193 \\ \hline
Overall             &$0.495 \pm 0.037$&$0.503 \pm 0.036$&1876  \\ \hline
\end{tabular}}
\caption{Notes: Results are given for the preferred \velmod\ runs:
300 \kms-smoothed \iras\ predicted velocity field;
density-dependent velocity dispersion with $\sigma_{v,1}=130\ \kms$
(forward) and $\sigma_{v,1}=110\ \kms$ (inverse); Paper I quadrupole
with $R_Q=3500\ \kms$, and corresponding value of $\bfwlg$.
%The likelihoods which yield 
The subsample $\beta_I$'s
were calculated using the
TF and velocity parameters obtained from
the full-sample run; these parameters were not solved for separately
for each subsample.}
%. Separate \velmod\ runs restricted to the
%subsamples above were not carried out.}
\label{tab:breakdown}
\end{table}
We test the robustness of our results by 
computing the maximum likelihood $\beta_I$ for different
TF subsamples and redshift ranges.
The is done in Table~\ref{tab:breakdown}
for our favored forward and inverse runs. Each of the four TF
subsamples, for both the forward and
inverse TF relations, produces a maximum likelihood $\beta_I$ consistent
with one another and with the global value of $0.50.$ Similarly, the
maximum likelihood $\beta_I$ for objects in each of five redshift bins
are statistically consistent with one another.  The last redshift bin
gives a value of $\beta_I$ somewhat higher than the others, but the
error bar is larger, and it is still consistent. 
%is somewhat discrepant, in the case of the inverse relation, but
%the error for that bin is quite large.  
Thus, there is no significant trend with redshift.
This consistency among sample and redshift
range enhances our confidence in our global value of $\beta_I.$
%{\bf It is not clear to me what was varied to give the numbers in the
%table.  Did you let the TF parameters vary, or did you simply add the
%individual likelihoods for the best TF parameters at each beta?}
%[JW - does my Table note (b) answer the question?
Note that lower-redshift galaxies give more leverage per object
on $\beta_I$ than do higher redshift galaxies. The reasons for
this were discussed in Paper I, \S4.5. The W91PP sample yields
the weakest constraints on $\beta_I,$ because of the relatively small
volume it probes.  Although there are fewer CF than
W91PP galaxies, they yield a stronger constraint on $\beta_I$ because
the CF sample has wider sky coverage. 

\subsection{Results for 500 \kms\ smoothing}
\label{sec:500-smooth}
%An important conclusion from 
Our Paper I tests with mock TF
and \iras\ catalogs showed that \velmod\ returned unbiased estimates
of $\beta_I$ when a 300 \kms\ Gaussian smoothing scale was used in the \iras\ 
velocity predictions. We also tested a
500 \kms\ smoothing scale and found that it produced
estimates of $\beta_I$ biased \sm 25\% high. For the
real data, 500 \kms\ smoothing produced a
maximum likelihood $\beta_I$ about 15\% higher
than the 300 \kms\ value (cf.\ Paper I, \S~4.6). 

\begin{figure}[th!]
\plottwo{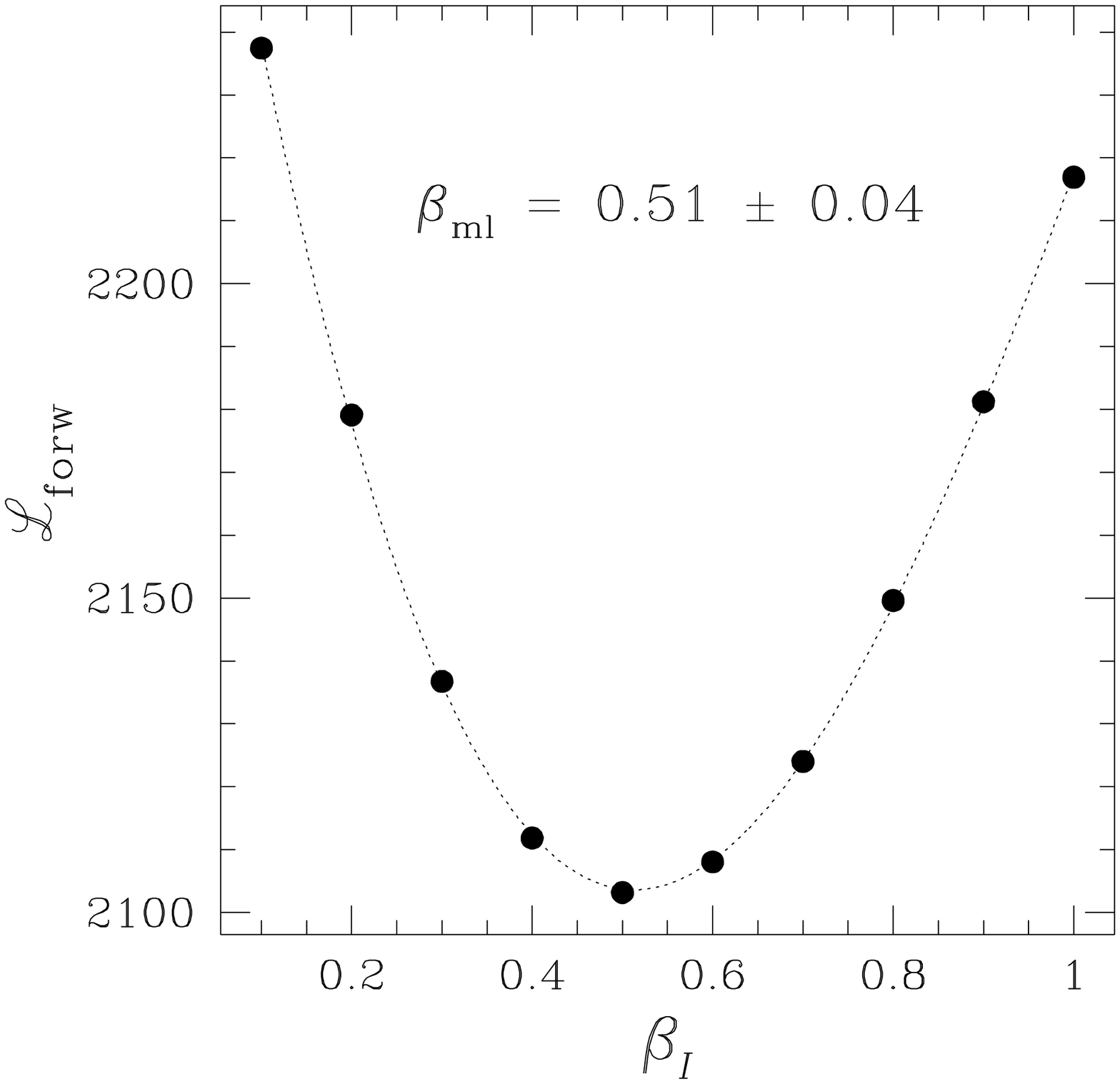}{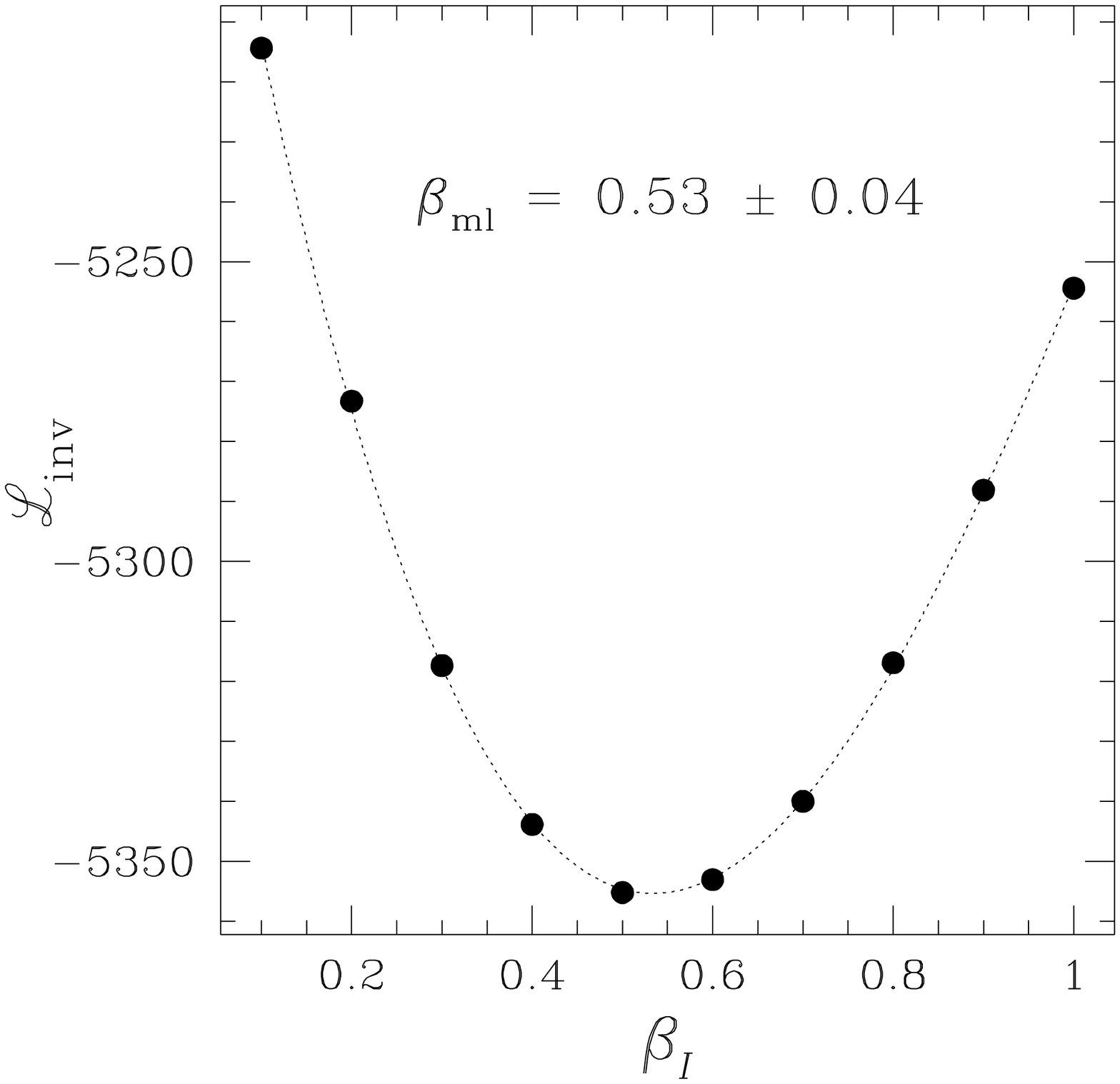}
\caption{{\small As in the previous figure,
now using the 500 \kms-smoothed \iras\ velocity
field predictions.  The maximum
likelihood values of $\beta_I$ 
%(indicated as $\beta_{{\rm ml}}$) 
differ by very little
from those of the previous figure.
%, in which the
%\iras\ predictions were done at 300 \kms\ smoothing.
}}
\label{fig:iras500}
\end{figure}
The results of applying \velmod\ to the expanded sample
using 500 \kms-smoothed \iras\ velocity predictions
are shown in Figure~\ref{fig:iras500}, in which 
likelihood for the forward and inverse
TF relations is plotted versus $\beta_I.$ These
runs are carried out using the values of the quadrupole parameters and
$\bfwlg$ obtained from the Paper I 500 \kms\ run, but again using the
modified quadrupole of equation~(\ref{eq:newquad}) with $R_Q=3500\ \kms.$
We again allow for a density-dependent value of $\sigma_v;$ we
show the results for $\sigma_{v,1}=150\ \kms,$ which maximizes
likelihood at 500 \kms\ smoothing. 

The 500 \kms\ maximum likelihood estimates of $\beta_I$
differ little from those obtained at 300 \kms\ smoothing.
Averaging the forward and inverse results, we find $\beta_I=0.52 \pm
0.05$, only 4\% higher than our 300 \kms\ result. 
%(the slightly enlarged
%error estimate incorporates the difference between forward and inverse
%in quadrature with the statistical errorbar).
In the \velmod\ analysis, the TF data are not smoothed, and therefore
we chose a small smoothing scale for the \iras\ density field in order
to model the velocity field in as much detail as possible.  The fact
that $\beta_I$, and, more significantly, the best
values of $\likeforw$ and $\likeinv,$ are essentially
unchanged when the smoothing is increased to 500
\kms, says that density fluctuations on scales between 300 and 500
\kms\ contribute little to the velocity field.  That is, there is
little small-scale power both in the true velocity field, and in the
\iras-predicted velocity field (i.e., the gravity field). 
%which is qualitatively
%consistent with our conclusion above that $\sigv$, which measures the
%components of the velocity field on scales smaller than 300 \kms, is
%small. [JW] This contradicts our mock catalog result where sigv was <100 km/s.
The simulations used in the mock catalogs in Paper I do have a
substantial amount of small-scale power in the density field, and this
is presumably the 
reason that they yielded a substantially biased estimate of $\beta_I$,
and a far worse value of $\like$, with 500 \kms\ smoothing. 
Because of the lack of small-scale power in the velocity field,
the agreement between our 300 and 500 \kms\ results
for $\beta_I$ does not shed light on the question
of whether biasing is scale-dependent.

Our 500 \kms\ runs also detect an
increase in the small-scale velocity dispersion with density. For
the forward run we find $f_\delta=30\ \kms,$ similar to
the $300\ \kms$ value. However, for
the inverse run we find $f_\delta=60\ \kms.$ We would expect
$f_\delta$ to rise with smoothing scale because the density contrasts
are generally smaller with larger smoothing. The inverse result
confirms this expectation but the forward does not; we do not
understand the reason for this difference.

\section{\velmod\ versus Mark III Catalog TF Calibrations}
\label{sec:calib}
\begin{table}[t]
\centerline{\begin{tabular}{l | c r c c c c c}
\multicolumn{8}{c}{TABLE 4: \velmod\ and Mark III TF Relations$^a$} \\ \hline\hline
 & \multicolumn{3}{c}{forward} & & \multicolumn{3}{c}{inverse} \\ \cline{2-4} \cline{6-8}
Sample & $A$ &\multicolumn{1}{c}{$b$}& $\sigtf$ & & $D$ & $e$ & $\sigeta$ \\ \hline
A82 (\velmod)      &$-5.96$ &$10.44$& 0.45 & &$-5.96$&$0.0879$& 0.042 \\ 
A82 (Mark III)     &$-5.94$ &$10.29$& 0.47 & &$-5.98$&$0.0893$& 0.043 \\ 
MAT (\velmod)      &$-5.80$ & $7.16$& 0.43 & &$-6.00$&$0.1282$& 0.063 \\
MAT (Mark III)     &$-5.79$ & $6.80$& 0.43 & &$-5.96$&$0.1328$& 0.059 \\ 
W91PP (\velmod)    &$-4.09$ & $7.14$& 0.40 & &$-4.13$&$0.1217$& 0.052 \\
W91PP (Mark III)   &$-4.28$ & $7.12$& 0.38 & &$-4.32$&$0.1244$& 0.049 \\ 
CF (\velmod)       &$-4.00$ & $8.41$& 0.48 & &$-3.97$&$0.0948$& 0.049 \\
CF (Mark III)      &$-4.22$ & $7.73$& 0.38 & &$-4.27$&$0.1190$& 0.047 \\ \hline
\end{tabular}}
\caption{Notes: (a) Comparison of the \velmod\ and Mark III TF calibrations for
the four subsamples used in the \velmod\ analysis. The typical \onesigma\ errors
for both calibrations are: $\delta A = \delta D \simeq 0.03;$
$\delta_b/b = \delta e/e = 0.03;$ $\delta \sigtf = e^{-1} \delta\sigeta = 0.02.$}
\label{tab:calib}
\end{table}

An important feature of \velmod\ is that the
TF relations for the various samples 
are determined by 
maximizing likelihood at each $\beta_I.$ 
The correct TF relations are those obtained
for the maximum likelihood value
of $\beta_I,$ which we have found to be $0.50$ with 
small uncertainty.
In Table~\ref{tab:calib} we give the parameters of the
forward and inverse TF relations obtained from our favored
\velmod\ runs, i.e.\ 300 \kms\ smoothing, quadrupole with
$R_Q=3500\ \kms,$ and density-dependent velocity dispersion
with $\sigma_{v,1}=130\ \kms$ (forward) and
$\sigma_{v,1}=110\ \kms$ (inverse).
Columns (1), (2), and (3) list the forward TF parameters
$A,$ $b,$ and $\sigtf,$ 
while columns (4), (5), and (6)
list the inverse TF parameters $D,$ $e,$ and $\sigeta$\footnote{The
scatters are given for $\eta=0$ and for the mean  
absolute magnitude in each sample; cf.\ \S~\ref{sec:TFscatter}.}.  Also given
in Table 3 are
the values of these parameters that went into the
Mark III Catalog (Willick \etal\ 1997a). 

The slopes and scatters of the \velmod\ and Mark III TF relations are
in good agreement overall. The \velmod\ MAT TF slope is higher, but by less
than $2\,\sigma,$ as we discussed in Paper I, \S 4.7. 
The CF slope is higher than its Mark III value, as is its scatter.
This is not surprising, because in this paper we have treated CF
as a fully independent sample, whereas the Mark III calibration
procedure (Willick \etal\ 1996) 
assumed that CF had the same TF relation as the Willick (1991)
cluster sample, W91CL, 
up to a slight zero-point adjustment. 

More important, there are substantial zero-point differences between
the \velmod\ and Mark III calibrations. 
While the \velmod\
and Mark III TF zero points of A82 and MAT are in good agreement, those
of W91PP and CF differ by about 0.2 mag, for both the forward and
inverse TF relations. 
This difference
is much greater than
the expected errors of \sm 0.03 mag in either procedure. 
Because the difference manifests itself
for only two of four samples, it cannot arise from a global zero point
error in either the Mark III or the \velmod\ calibration procedure.

\begin{figure}[t!]
\plottwo{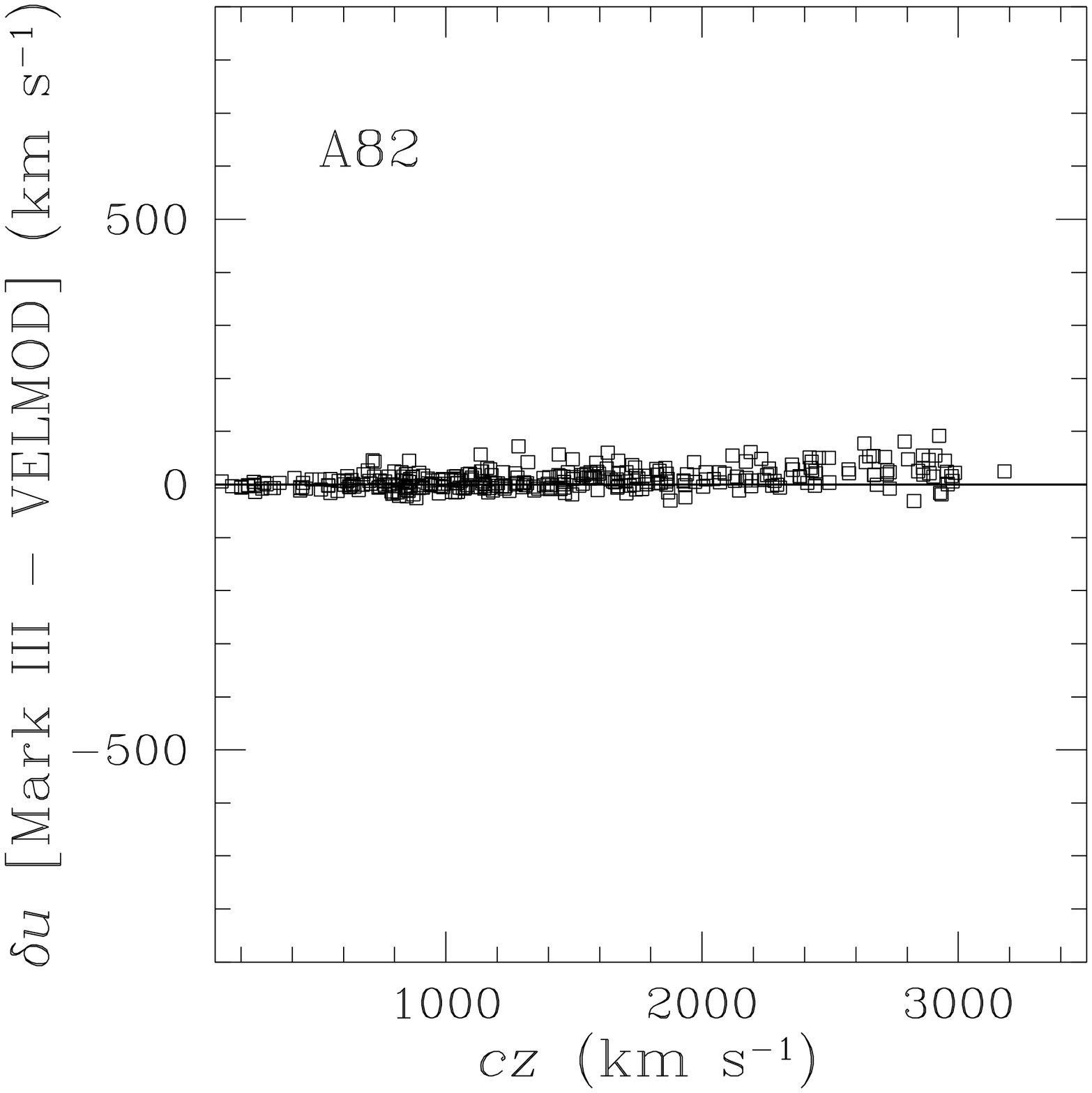}{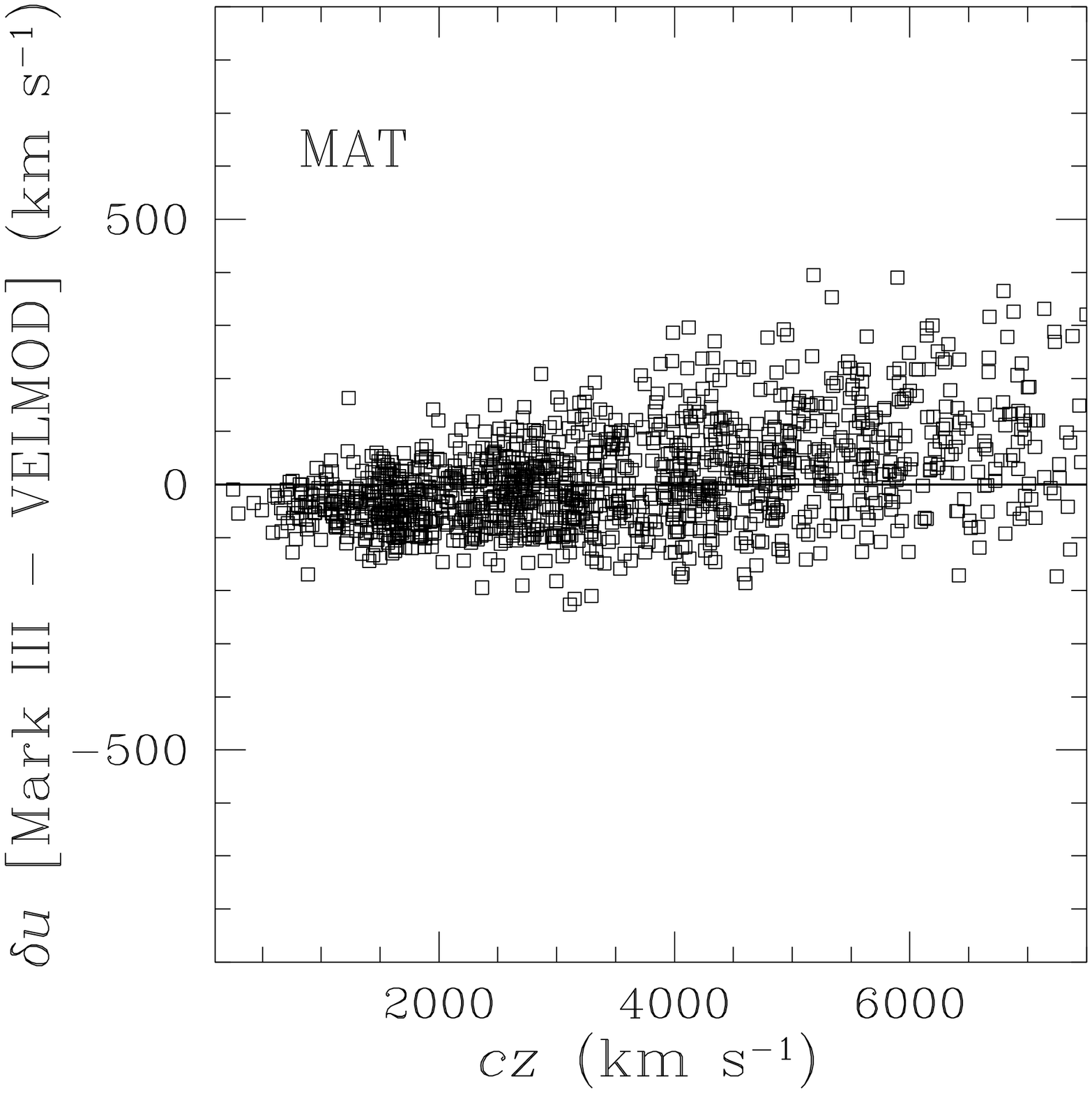}
\caption{{\small Differences
in the radial peculiar velocity inferred from the
Mark III and \velmod\ forward TF calibrations,
plotted as a function of Local Group redshift for
the A82 (left panel) and MAT (right panel) samples.
}}
\label{fig:amcomp}
\end{figure}

\begin{figure}[t!]
\plottwo{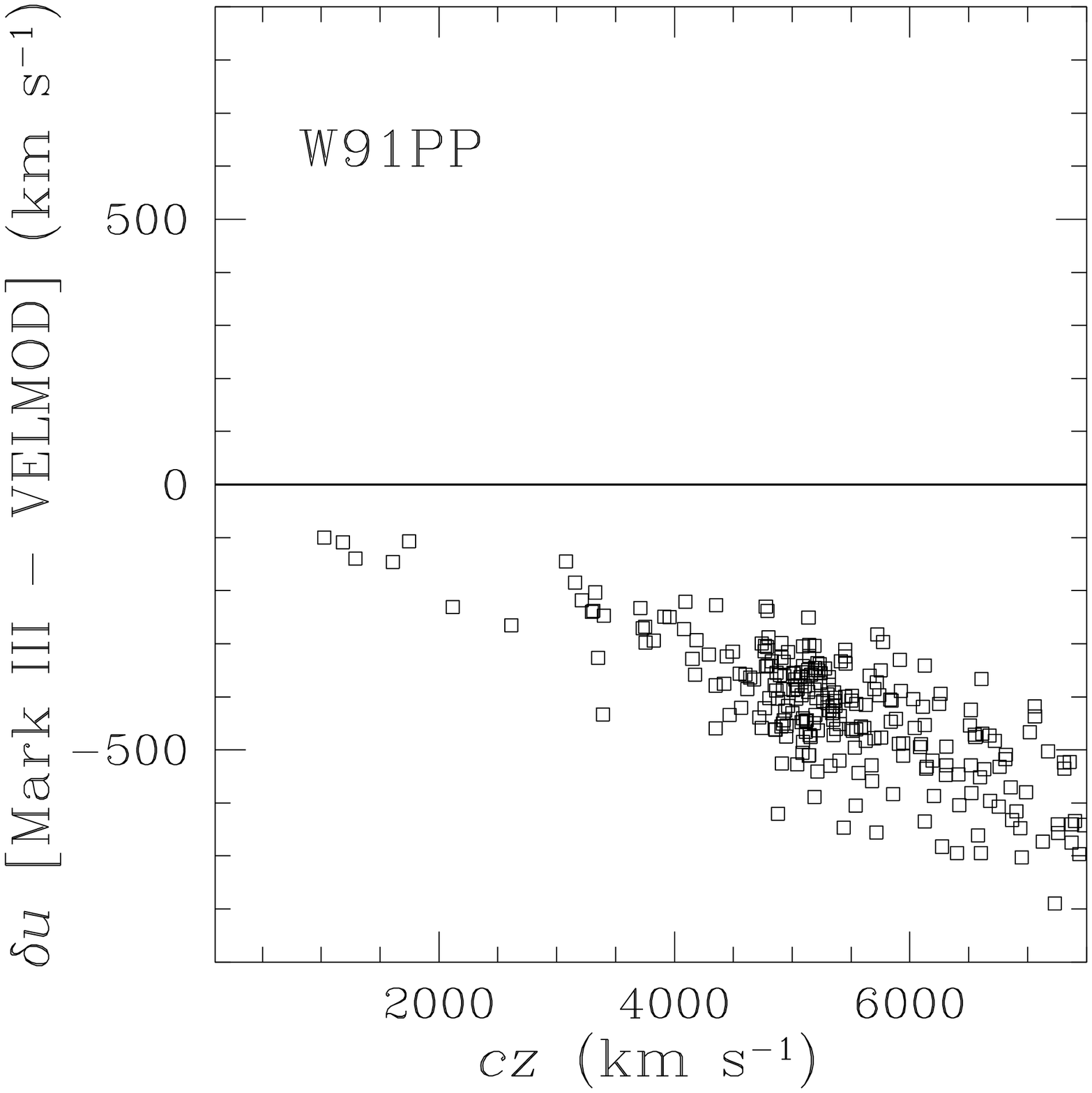}{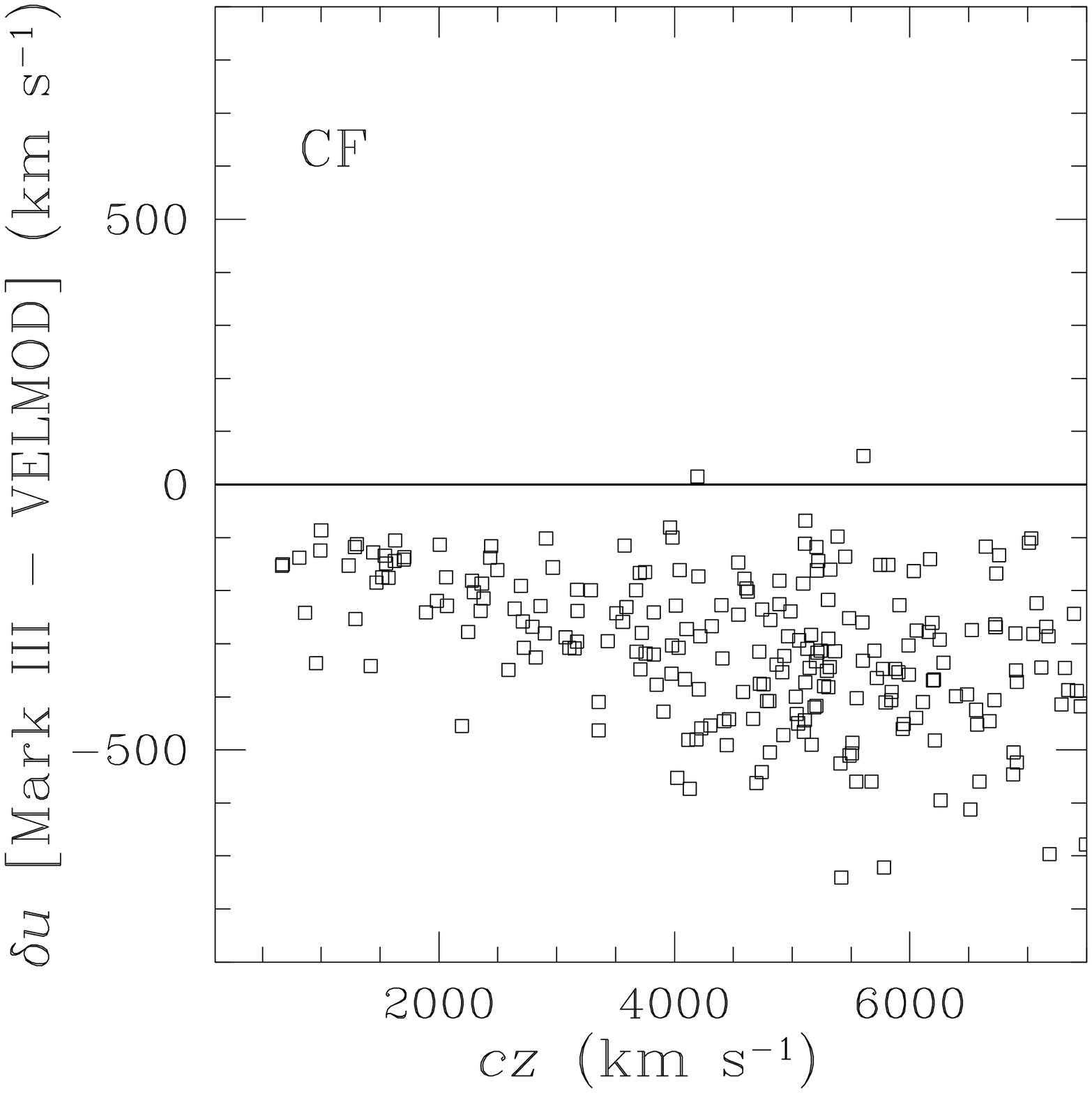}
\caption{{\small As in the previous figure,
for the W91PP (left panel) and CF (right panel) samples.}}
\label{fig:wcfcomp}
\end{figure}

Figures~\ref{fig:amcomp} and~\ref{fig:wcfcomp} show how
these differences in the TF parameters translate into 
peculiar velocity differences.
The differences between the Mark III and \velmod\ peculiar
velocities inferred from the forward TF relation
are plotted
as a function of LG redshift for each of the
four samples. The plots would
appear substantially the same if we used inverse TF distances.
We do not apply
Malmquist bias corrections, 
which would accentuate the differences
between the \velmod\ and Mark III velocities. Thus, the
TF scatters have no effect on the diagrams.

For A82 there is
no meaningful difference between
the Mark III and \velmod\ inferred peculiar velocities. For MAT,
there is a slight trend, but the mean differences are everywhere
less than \sm 100 \kms, except at the outer edge ($cz \simeq 6000\ \kms$)
of the sample. 
However, for W91PP and CF the differences 
are substantial.
In each case, the Mark III velocities are more
negative by 200--400 \kms. In the case of W91PP, the differences
are even larger 
beyond 6000 \kms.\footnote{For W91PP the trend is essentially
linear with redshift, and has small scatter, whereas for CF, there is larger scatter
and the velocity difference levels off at large redshift. This
is because for W91PP the calibration difference involves only
the TF zero point, while for CF both zero point and slope
differences are present. The TF slope difference also explains why the
MAT diagram exhibits a much larger scatter than the A82 diagram.}

This systematic difference between the Mark III and
\velmod\ TF calibrations has a strong effect on the inferred bulk flow
from the Mark III data (cf.\ Courteau \etal\ 1993; Dekel 1994; Postman
1995; Strauss 1997, for discussions).  
The W91PP and CF samples
dominate the Northern sky 
away from the Local Supercluster. W91PP in particular samples the
Perseus-Pisces (PP) 
supercluster, centered at $l\approx 120\degs,$ $b\approx -30\degs.$ As
measured by the Mark III TF calibrations, the PP
region is seen as having large, negative radial peculiar velocities
in the microwave background frame (e.g., 
Courteau \etal\ 1993). 
This, along with outflowing velocities in the Great Attractor
region (traced mainly by the MAT sample), is why measurements of the
bulk flow within 6000\ \kms\
from the Mark III data have yielded values in the range $400$--$500\ \kms$.
However, \iras\ does
not predict strong infall of the PP supercluster region, unless
$\beta_I$ is $\simlt 0.2.$ 
Since the \velmod\ TF calibrations reflect the \iras\ velocity field,
they adjust to produce little infall of PP, and thus a much smaller
bulk flow, than do the Mark III calibrations. 

Another way to state the problem is as follows. 
The Mark III TF zero points were set by asking for
agreement in distances for galaxies in overlapping datasets; the
full-sky cluster sample of Han \& Mould (1992; HMCL) was the backbone that
tied the sky together (cf.\ Willick \etal\ 1995, 1996, 1997a).  If
these calibrations are indeed 
correct, then the \velmod\ calibrations are not, and
it follows that the \iras\
redshift survey does not correctly predict the peculiar velocity field.
In fact, this was the conclusion reached by Davis \etal\ (1996),
whose ITF analysis made use of the Mark III 
zeropointing procedure even though it did not use the Mark III distances directly.  
If the \iras\ velocity field predictions are correct,
as we have assumed in this paper, then so are the \velmod\
TF calibrations and our maximum likelihood estimate
of $\beta_I.$   However, in that case the Mark III
TF calibrations are incorrect, and the Mark III Catalog contains erroneous
distances for the W91PP and CF samples---and
by extension, for the HMCL, W91CL, and elliptical
galaxy samples as well. 
It would then follow that the POTENT peculiar velocity
and density fields, which are based on the Mark III distances
and were used in the POTIRAS determination
of $\beta_I=0.89\pm 0.12,$ 
contain systematic errors. 
A self-consistent picture would
require that the \velmod\ TF calibrations, required by the
\iras\ velocity fields, also be used to produce the POTENT
velocity and density maps to estimate $\beta_I.$ This
has not yet been done.

One can ask whether the \velmod\ 
TF calibrations agree better with
the Mark III calibrations for some value of $\beta_I$ other than $0.5.$
In fact, for $\beta_I = 0.1$, the \velmod\ W91PP zero point agrees with
that of Mark III, for both the forward and inverse TF relations. 
However, for $\beta_I=0.1$
the \velmod\ TF zero point for CF is even farther from
its Mark III value than it is for $\beta_I=0.5.$ For $\beta_I\simeq 1,$
the CF zero point is closer to its Mark III value, but the W91PP zero
point diverges drastically from Mark III. Also, for very low or very
high $\beta_I$ we lose the good agreement between the \velmod\ 
and Mark III A82 and MAT TF zero points.
Thus there is no value of $\beta_I$ at which the \velmod\ and Mark
III calibrations are in overall agreement. 

The question of which set of
TF calibrations is correct must ultimately be
decided by improved TF data. The
problem has arisen because there is no reliable way
to tie together the disjoint
Southern (MAT) and Northern (CF and W91PP) sky TF data sets
that constitute the Mark III field spirals. 
A82 spans the two hemispheres but is dominated by
nearby 
galaxies and has little overlap with the Northern sky samples. 
The HMCL sample was thought to provide the needed overlap, but its uniformity
across the sky has been called into question by the calibration disrepancies.
What is needed are homogeneous TF data that cover the celestial
sphere. 
In collaboration with S. Courteau, M. Postman,
and D. Schlegel, we have  
obtained uniform TF data for \sm 300
galaxies isotropically distributed in the spherical shell defined by 
$4500 \simlt cz \simlt 7000\ \kms.$ Reduction of these data are
under way, and results are expected by late 1998. Comparison of
these uniform TF data with the Mark III data will allow a definitive
resolution of the calibration problem.

Finally, we note that adopting the Mark III
TF calibrations has relatively little effect on
the maximum likelihood $\beta_I$ obtained
from \velmod. With the $300\ \kms$-smoothed \iras\ plus quadrupole
velocity model, we obtain $\beta_I=0.44$ (forward)
and $\beta_I=0.45$ (inverse) when the TF parameters for
all four samples are fixed to their Mark III values
as given in Table~\ref{tab:calib}. For the no-quadrupole
model we obtain $\beta_I=0.50$ (forward) and $\beta_I=0.51$
(inverse). The likelihoods obtained from these
\velmod\ runs are, of course, much worse (by \sm 100
units in $\like$) than for our preferred runs in
which the twelve TF parameters are free. Thus, while the TF calibration
problem is crucial for the match of the
\iras\ velocity field to the TF data, as we discuss
in the next section,
it is secondary for the determination of $\beta_I.$

\section{The Goodness of Fit of the \iras\ Velocity Field}
\label{sec:residuals}
Although \velmod\ does not produce 
a picture of the TF velocity field, we can nonetheless 
use it to visualize how well the
TF data fit the \iras\ velocity predictions.   
We do so by converting
the \velmod\ apparent magnitude $m$ (forward) or velocity width
parameter $\eta$ (inverse)
residuals into smoothed radial peculiar velocity residuals with respect to
\iras, as described in
Paper I, \S~5.1. The \velmod\ residuals also enable us
to measure the goodness of fit of the velocity model, as we describe below.
The smoothed peculiar velocity residual is given by
equation~24 of Paper I:
\begin{equation} 
\delta u_i^s = d_i \left[1-f_i 10^{0.2(\delta_{m,i}^s\times \Delta
m_i)}\right]\,,
\label{eq:smooresid}
\end{equation}
which we repeat here because of a typographical error in Paper I; see
Paper I for the definition of the various symbols in this
equation. 

\begin{figure}[t!]
\centerline{\epsfxsize=7.5 in \epsfbox{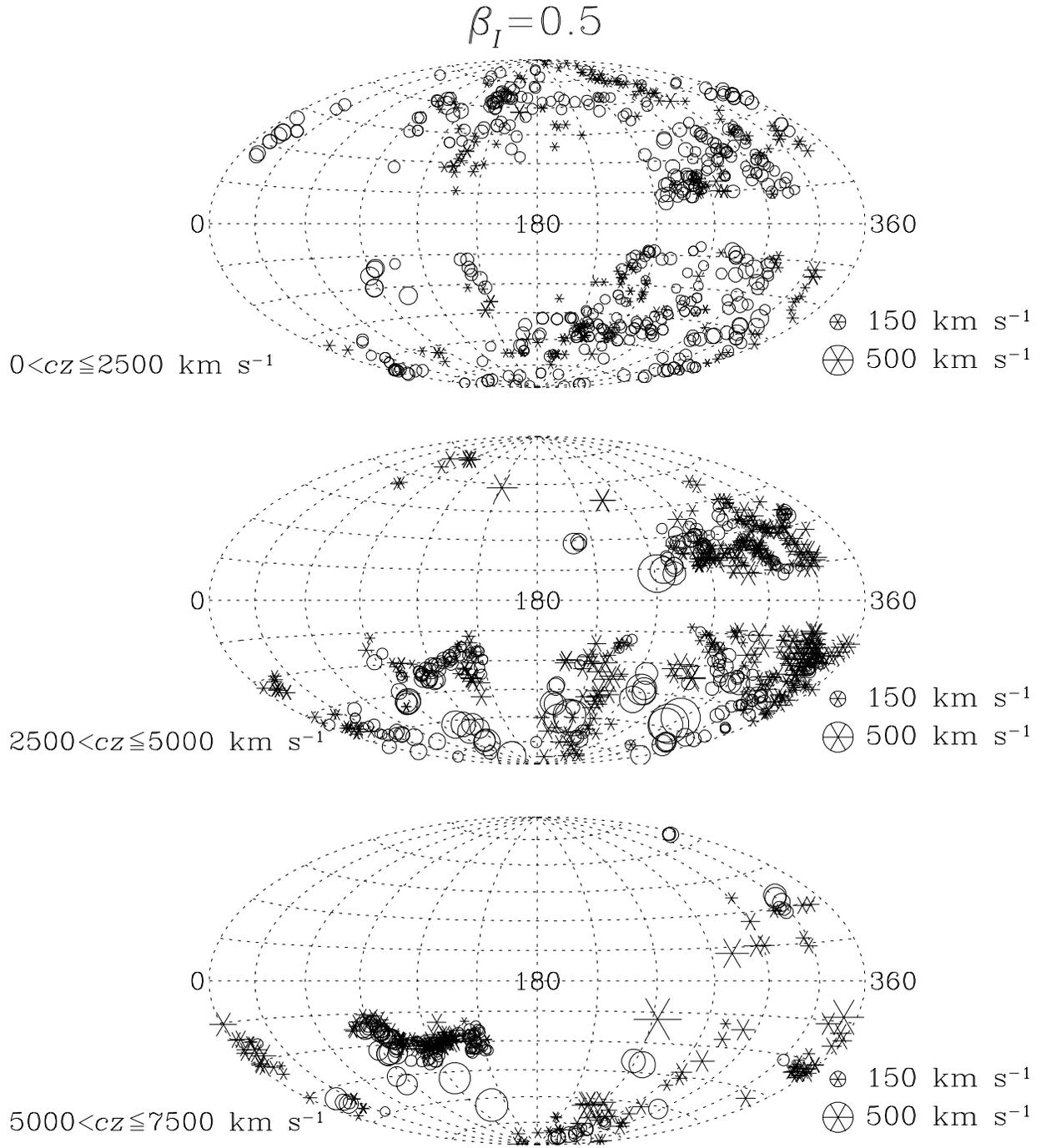}}
\caption{{\small Smoothed \velmod\ velocity residuals plotted
in Galactic coordinates, for $\beta_I=0.5.$
Open circles indicate objects
inflowing relative to the velocity model, while starred
symbols represent outflowing objects. The symbol size
indicates the magnitude of the velocity residual,
as coded at the lower right of each plot.}}
\label{fig:residmap5}
\end{figure}

\begin{figure}[t!]
\centerline{\epsfxsize=7.5 in \epsfbox{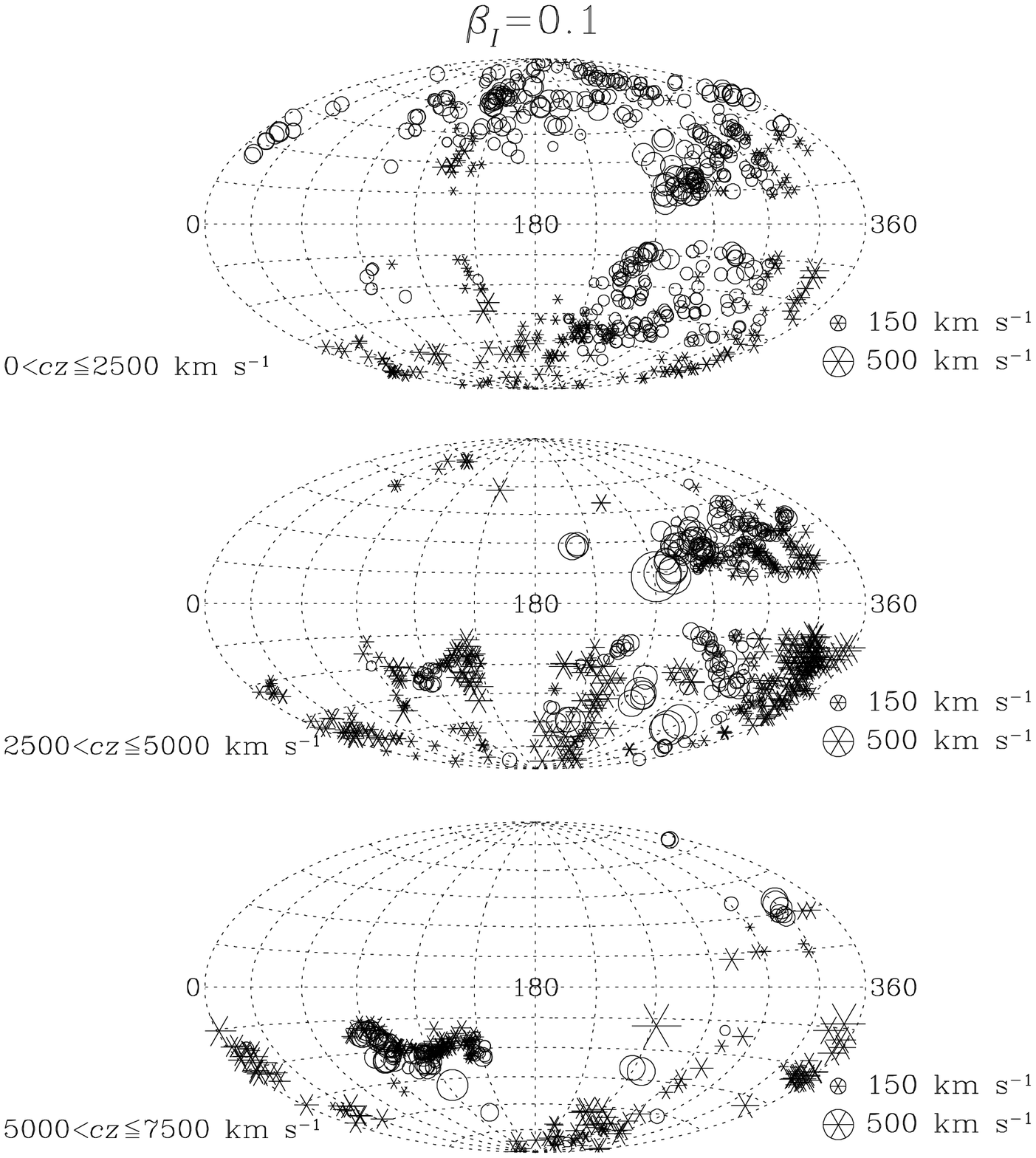}}
\caption{{\small Same as the previous figure, but for $\beta_I=0.1.$}}
\label{fig:residmap1}
\end{figure}

\begin{figure}[t!]
\centerline{\epsfxsize=7.5 in \epsfbox{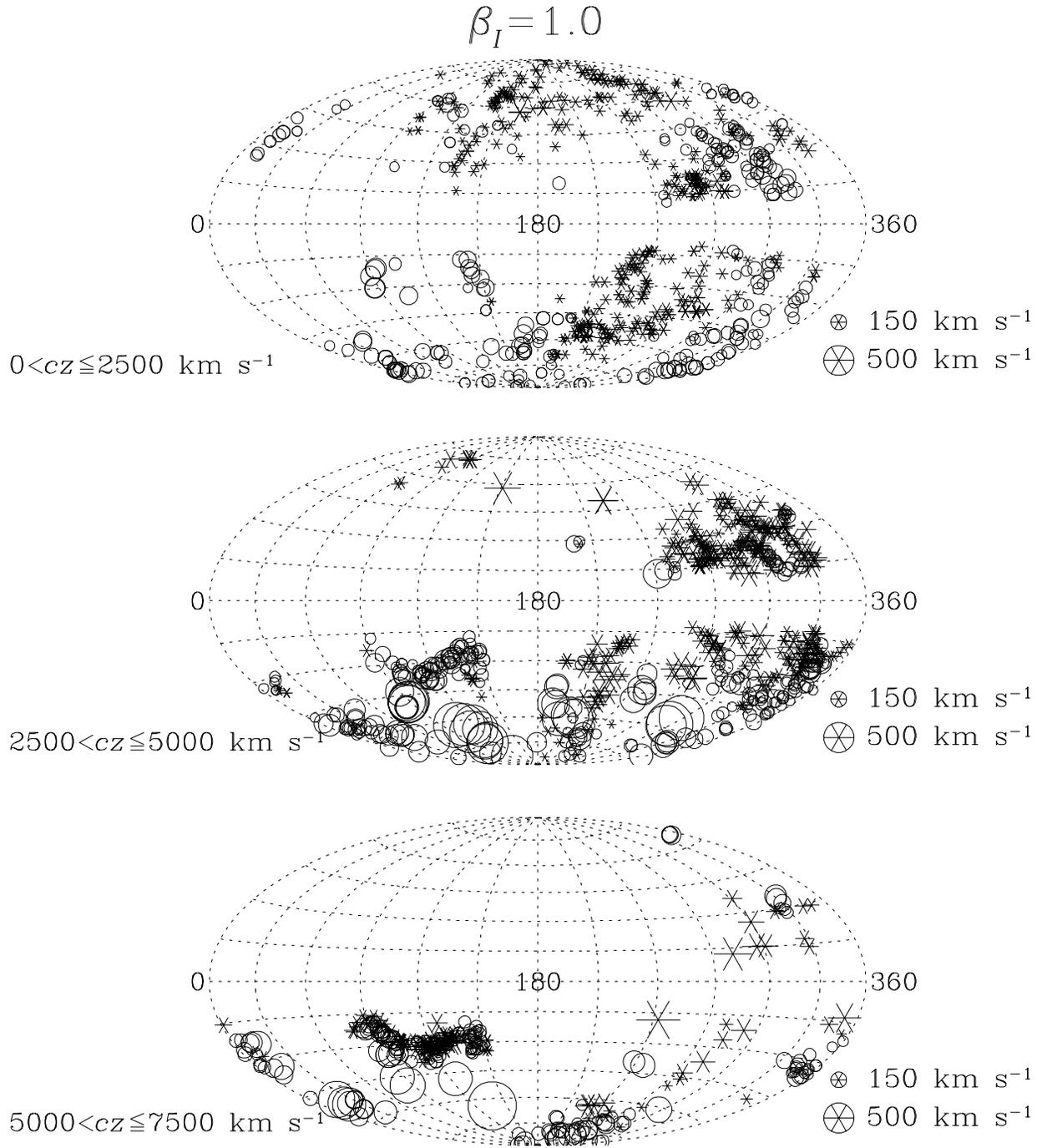}}
\caption{{\small Same as the previous figure, but for $\beta_I=1.0.$}}
\label{fig:residmap10}
\end{figure}

Figures~\ref{fig:residmap5}, \ref{fig:residmap1}, and~\ref{fig:residmap10}
show sky maps of these velocity residuals for $\beta_I=0.5,$ $\beta_I=0.1,$
and $\beta_I=1.0$ respectively. In each case, the results are based on 
forward TF residuals from our
preferred 300 \kms\ smoothing run (see the notes to Table 3).
Open symbols represent negative velocity residuals (i.e., the TF distance
to the object is greater than that predicted by \iras); starred symbols
represent positive velocity residuals.
The Gaussian smoothing scale for the maps is given by 
%$r_{{\rm smoo}}=
$250\left[1+(\czlg/2500)^2\right]^{1/2}\ \kms.$ 
Thus, the smoothing radius varies
from 250 \kms\ nearby to \sm 750 \kms\ at the edge of the sample.
This smoothing imposes a coherence scale of \sm 15--25\degs\
on the results; patches this size with similar velocity residuals
are to be expected in the maps from the smoothing alone, while any coherence
seen on much larger scales represents a real error in the model. 
Points are plotted only for galaxies which have enough near
neighbors to allow an adequate smoothing; this is 
%whose effective weight
%from the smoothing procedure is 4 or greater at $cz\leq 5000\ \kms$
%and 5 or greater at $cz>5000\ \kms.$ 
%[MS: Are you happy with my rewording?  This avoids having to give a
%proper definition of ``weight''.]
%This explains 
why there are few
galaxies represented in the Northern Galactic Cap at $cz>2500\ \kms,$
where the sampling is very dilute. Such points, if plotted, would
exhibit large velocity residuals due solely to TF scatter and
would not help us assess the quality of the fit.
%{\bf I am struck by the absence of galaxies in the Northern Galactic
%Cap at $cz > 5000 \kms$; is this due to the way CF has been cut?  It
%may be worth mentioning how good it would be to get the SFI sample in
%here!}

Inspection of these maps shows clearly why $\beta_I=0.5$ is the best fit.
Although there is some real excess coherence to the residuals (we discuss
this further below), the coherent velocity levels are generally at a low
level ($\simlt 250\ \kms$). There are many alternating regions of positive
and negative residuals, showing that globally at least the residual map
is fairly incoherent. This is what is required of a good fit. There are no
extended regions where the velocity residuals are consistently greater than
300 \kms.  This is a qualitative indication that the \iras\ plus
quadrupole velocity field
model fits the major features of the actual velocity field.  
As in Paper I,
coherent residuals are present when the quadrupole is not modeled.
That being said, with $R_Q
= 3500\ \kms$, the quadrupole contribution at $cz \simgt  5000\ 
\kms$ is negligible. Thus, the good agreement on very large
scales is due to the \iras\ velocity field alone, giving \aposteriori\
confirmation of our quadrupole model.  

The residual maps produced at $\beta_I=0.1$ and at $\beta_I=1.0,$ on the other
hand, show considerable coherence. Moreover, the amplitude of the
velocity residuals in these regions is often large, $\simgt 300\
\kms.$ Low and high $\beta_I$ are clearly worse fits to the TF data
than is $\beta_I=0.5.$ The maps, then, confirm what the likelihood
analysis is telling us. It is important to remember that the poor fit
at low and high $\beta_I$ is not a result of errors in the assumed TF
relation, for the TF relations used were those preferred by the data
at each $\beta_I$. The poor fit is a genuine reflection of the
incorrectness of the \iras\ velocity field for low and high $\beta_I.$

\clearpage

\clearpage

\clearpage

\begin{figure}[t!]
\centerline{\epsfxsize=4.5 in \epsfbox{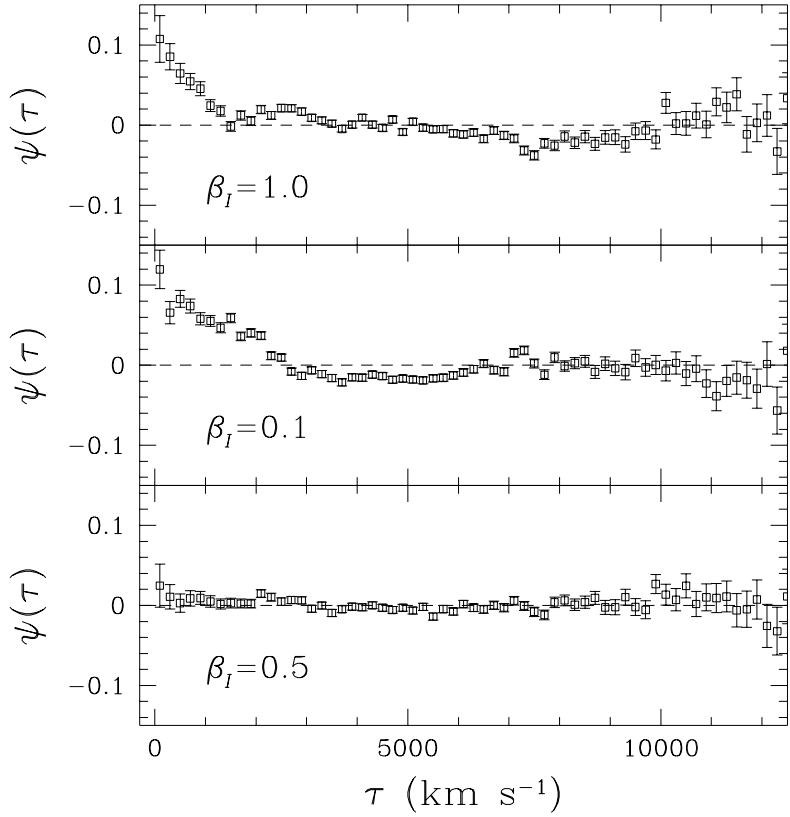}}
\caption{{\small \velmod\ residual autocorrelations, $\psi(\tau)$, plotted
for $\beta_I=0.5,$ $\beta_I=0.1,$ and $\beta_I=1.0.$ Nonzero values of
$\psi(\tau)$ indicate coherence of the TF versus \iras\ residuals on
a spatial scale $\tau.$ Such coherence is pronounced for $\beta_I=0.1$
and $\beta_I=1.0,$ indicating a poor fit. It is insignificant for $\beta_I=0.5.$}}
\label{fig:autocorr}
\end{figure}
We may quantify our visual impressions by means of
the residual autocorrelation function $\psi(\tau),$ defined
by equation~(25) of Paper I. 
In Figure~\ref{fig:autocorr},
we plot $\psi(\tau)$ for the three values of $\beta_I$ represented
in the previous figures. The plots show that for $\beta_I=0.1$
and $\beta_I=1.0,$ significant excess correlation is evident on small and large
scales. At $\beta_I=0.5,$ the $\psi(\tau)$ is consistent with
zero on all scales. There is a small amount of positive correlation
on scales $\simlt 2500\ \kms$ for $\beta_I=0.5,$ consistent with the
(low-amplitude) inflowing monopole residuals in the upper panel of
Figure~\ref{fig:residmap5}. This may be 
indicative of a breakdown
of the \iras\ model at some level, but it is not highly significant,
as we now show.

\begin{figure}[t!]
\centerline{\epsfxsize=4.5 in \epsfbox{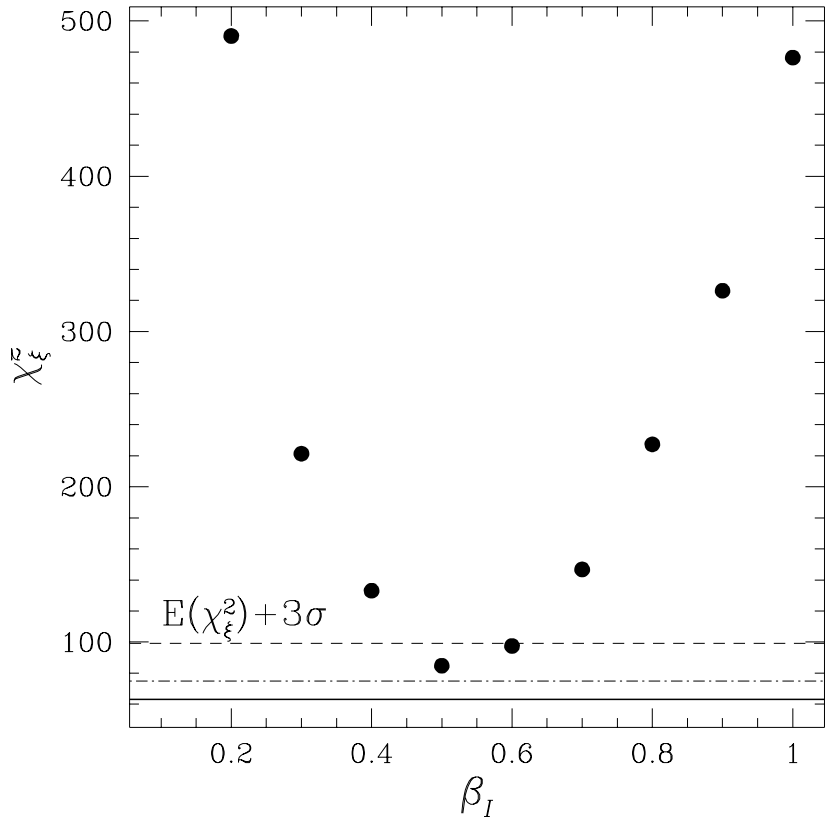}}
\caption{{\small The residual correlation statistic $\chi^2_\xi,$ plotted
as a function of $\beta_I.$ The heavy line shows the expectation value
of this statistic; the dot-dashed and dashed lines show the \onesigma\
and $3\,\sigma$ deviations respectively.}}
\label{fig:chi2xi}
\end{figure}
A rigorous measure of the level of residual coherence comes through the
use of the correlation $\chi^2$ statistic, $\chi^2_\xi,$ defined by
equation~(26) of Paper I. 
We plot $\chi^2_\xi$ versus $\beta_I$ in Figure~\ref{fig:chi2xi}. (The
value for $\beta_I=0.1$ is off-scale.)
%is not shown because it is so large.)
In Paper I we showed that this statistic had properties similar to
that of a true $\chi^2$ statistic, but with a mean of $0.87 \pm 0.06$
per degree of freedom rather than unity. Its variance was consistent
with that of a true $\chi^2$ statistic.  We indicate the expected
value of $\chi^2_\xi$ (in this case, 63.5 for 73 degrees of freedom)
as a heavy solid line on the plot. The 1- and $3\,\sigma$
deviations from the expectation are indicated as dot-dashed and
dashed lines, respectively.  The quantity $\chi^2_\xi$ reaches its
minimum at the maximum likelihood value of $\beta_I.$ The only other
value of $\beta_I$ for which $\chi^2_\xi$ is within $3\sigma$ of the
expectation value is $\beta_I=0.6.$ $\beta_I\leq 0.4$ and $\beta_I\geq
0.7$ are ruled out at the $>3\,\sigma$ level.

%We note, finally, that the value of $\chi^2_\xi$ for $\beta_I=0.5$
%is $1.8\,\sigma$ above the expectation value. Thus, while acceptable,
%$\beta_I=0.5$ does exhibit larger residuals than might be expected for
%a perfect fit, as is indeed suggested visually by the sky maps.
%Either the \iras\ model, or the TF data, or both
%are imperfect at some level. However, the residual coherence 
%as measured by $\chi^2_\xi$ 
%is at a low enough level not to invalidate
%our conclusions about $\beta_I.$
%{\bf I'm not sure that the 1.8$\sigma$ deviation is worth getting
%worried about, given our lack of rigor of the value 0.87.  Indeed,
%there is absolutely no reason to believe that the same value should
%hold for this expanded sample.}

\section{Summary}
\label{sec:summary}
We have applied the \velmod\ method to a TF sample drawn from
the Mark III Catalog in order to estimate
$\beta_I=\Omega^{0.6}/b_I,$ where $b_I$ is the linear biasing
parameter for \iras\ galaxies. The TF sample consists of 1876
galaxies, comprising nearly all
Mark III field spirals to a limiting redshift of $\czlg = 7500\ \kms.$
This analysis extends the one we presented in Paper I,
which was limited to 838 galaxies with $\czlg\leq 3000\ \kms.$
As in Paper I, peculiar velocities were predicted
from galaxy density contrasts obtained from the \iras\ 1.2 Jy redshift
survey (Fisher \etal\ 1995), under the assumption of linear
gravitational instability theory and linear biasing. 
We developed an analytic approximation
to the single-object \velmod\ likelihoods, 
applicable to $\simgt 75\%$ of sample objects, which makes the code
run 3--4 times faster.

We carried out the \velmod\ analysis
using both the forward and inverse forms of the TF relation.
Consistency between the two is required to ensure that selection
biases are unimportant. We found that the maximum likelihood
values of $\beta_I,$ as well as other important velocity parameters,
were indeed statistically the same for both forms of the TF relation.
In addition, we allowed the quadrupole velocity residual detected
in Paper I to cut off smoothly beyond a radius, $R_Q,$ whose value
we determined through likelihood maximization to be
$3500 \pm 1000\ \kms.$  There is little covariance between $R_Q$ and
$\beta_I$. 
We believe the quadrupole is real and readily accounted for
(cf.\ Paper I, Appendix B).
%But further work to validate its existence
%is required.  
%[MS: Note that I have moved this sentence.  I took out the second
%clause, because it is not clear what further work you have in mind.]
We may summarize our results as
$\beta_I=0.50 \pm 0.04 \pm 0.04,$ where the first errorbar
is statistical and the second is systematic. This value is
quoted for our favored model in which the \iras\ densities
are smoothed with a 300 \kms\ Gaussian, the
small-scale velocity dispersion varies with density (see below), and in
which the quadrupole, with $R_Q=3500\ \kms,$ is
added to the \iras-predicted velocity field. The systematic
error is due to the quadrupole; if it is not valid to add it, 
%bar is associated with question of whether or not it
%is valid to add a quadrupole velocity residual. If it is not,
we obtain $\beta_I=0.53 \pm 0.04$ (forward), or 
$\beta_I=0.54\pm 0.04$ (inverse).
We also found that changing the \iras\ smoothing
scale from 300 to 500 \kms\ does not significantly affect
the derived value of $\beta_I.$ 
This implies that there is little contribution to the velocity and
gravity fields from fluctuations on scales between 300 and 500 \kms.
Further work needs to be done to quantify this, and to understand what
effect our Wiener filter, which suppresses fluctuations at large
distances, might have on this result. 

We tested for a density-dependence of the
small-scale velocity dispersion, by modeling
a linear variation of $\sigma_v$ 
with the galaxy density contrast
$\delta_g$ and determining the coefficient through
likelihood maximization. 
This significantly improved the \velmod\ likelihood, with a best fit
relation $\sigma_v = [(100 \pm 25) + 35\,\delta_g]\ \kms.$ This
confirms and strengthens our Paper I result that the galaxy velocity field
is remarkably cold.
Our detection of an increase in $\sigma_v$ with density agrees
qualitatively with
the results of Strauss \etal\ (1998), but our
coefficient of $\delta_g$ is considerably smaller than their
value of \sm 50--100 \kms.

We showed that the \iras-predicted velocity field, with quadrupole, is a good fit
to the TF data; the correlation function of velocity
residuals at $\beta_I=0.5$ is consistent with zero on all scales.
Strong velocity residual correlations on both small and
large scales are seen for
$\beta_I\leq 0.4$ and $\beta_I\geq 0.7,$ indicating
that the \iras-predicted velocity field is 
not a good fit for these values of $\beta_I.$
Davis \etal\ (1996), who adopted the Mark III TF zero points,
found highly significant discrepancies between the 
\iras-predicted and Mark III-observed velocity fields at all $\beta_I.$  
The \velmod\
procedure requires no \apriori\ calibration of the TF relation, and
with this freedom, the \iras-predicted velocity field matches the TF
data well, suggesting that the Davis \etal\ (1996) discrepancies are
tied to uncertainties in the TF calibrations. 
Our claim of agreement between the predicted and
observed velocity fields can hold up only if the
\velmod\ TF calibrations ultimately prove correct.

Indeed, we showed by direct comparison of TF parameters % (\S~\ref{sec:calib})
that the \velmod\ and Mark III TF calibrations (Willick \etal\ (1997a)
differ significantly.
The \velmod\ TF relations for CF and W91PP 
yield distances \sm 8\% shorter than the Mark III TF calibrations, whereas the
\velmod\ and Mark III TF calibrations for A82 and MAT are in good agreement.
This has a strong effect on the large-scale bulk flow inferred from
the data.
The \velmod\ TF calibrations cannot be brought into
closer agreement with the Mark III
calibration by changing $\beta_I,$ or by an overall zero point
shift in all TF samples. 
If the \velmod\ 
TF relations are correct, then the overall Mark III TF calibration
cannot be. Analyses based on the published Mark III
distances should thus be interpreted with caution. 

The \velmod\ TF calibrations are valid, however,
only to the degree that the \iras-predicted peculiar velocities are
accurate. This will be the case provided that \iras\ galaxies trace mass up
to linear biasing, and linear gravitational instability theory
is a good approximation when the galaxy densities are smoothed
on a 300--500 \kms\ Gaussian scale. Ultimately, the calibration
issue must be settled by improved observational data.  
We are carrying out a full-sky TF for this purpose,
and will report the results of
this effort in 1--2 years. 

Our result $\beta_I\simeq 0.5$ is virtually unchanged from % our
Paper I, %result, 
ruling out the possibility that cosmic scatter
and the small volume studied biased our Paper I $\beta_I.$ Thus,
this paper sharpens the discrepancy between the
\velmod\ measurement of $\beta_I$ and that obtained from
the POTIRAS comparison, $\beta_I\simeq 0.89 \pm 0.12$ (Sigad \etal\ 1998).
Further underscoring this discrepancy are
two analyses that have appeared since Paper I,
that of Riess \etal\ (1997) who find
$\beta_I=0.4\pm 0.15$ using SN Ia as tracers of the velocity field, and
that of da Costa \etal\ (1997) who found
$\beta_I=0.6\pm 0.12$ using the SFI TF data set. 
It may be that the
differences in the derived values of $\beta_I$ center on whether
the comparison is done at the level
of the velocities (the \vv\ comparison, as in this paper, Riess \etal,
and da Costa \etal) or at the level of the densities (the \dd\ comparison,
as in POTIRAS). Future work is needed to determine whether these
differences can be
explained in terms of physical effects, such as
a scale-dependent biasing relation (e.g., Sigad \etal\ 1998),
or whether they result from TF calibration errors, as discussed above,
or other methodological factors.
The question is
an important one because the values of
$\beta_I$ obtained from the \vv\ analyses
favor a low-density ($\Omega=0.2$--$0.5$) universe,
while the POTIRAS $\beta_I$ is suggestive of
an $\Omega=1$ cosmology, if 
%. In making these statements,
%we are assuming that 
$b_I \simlt 1,$ as suggested
by recent analyses of the evolution of rich clusters
(Bahcall, Fan, \& Cen 1997; Fan, Bahcall, \& Cen 1997). %; Bahcall 1997; Cen 1997).

\acknowledgements
JAW acknowledges the support of NSF grant AST-9617188.
MAS acknowledges the support of the Alfred P. Sloan Foundation,
Research Corporation, and NSF grant AST96-16901.  
We thank the members of the Mark III team, David Burstein,
St{\'e}phane Courteau, Avishai Dekel, and Sandra Faber, for their efforts
over the years in putting together the Mark III dataset.
We further thank St{\'e}phane Courteau for discussions concerning
selection of the CF sample, and Marc Davis and Tsafrir Kolatt for
comments on the text.

\appendix
\section{Appendix: Derivation of Approximate Likelihoods}
\label{sec:appendix}

The full expressions for the \velmod\ likelihoods are given by
equations~11 and 12 of Paper I:
\begin{equation}
P(m|\eta,cz)=\frac{\int_0^\infty\!dr\,r^2 n(r)\,P(cz|r)\,S(m,\eta,r) \expmM}
{\int_0^\infty\!dr\,r^2 n(r)\,P(cz|r)\,\infint\!dm\,S(m,\eta,r) \expmM}\,;
\label{eq:pmgetaz}
\end{equation}
\begin{equation} 
P(\eta|m,cz)= \frac{\int_0^\infty\!dr\,r^2 n(r)\,\Phi(m-\mu(r))\,P(cz|r)\,S(m,\eta,r)
\expet}
{\int_0^\infty\!dr\,r^2
n(r)\,\Phi(m-\mu(r))\,P(cz|r)\,\infint\!d\eta\,S(m,\eta,r) \expet}\,,
\label{eq:petagmz}
\end{equation}
where 
\begin{equation}
P(cz|r) = \nsigv\expcz\,,
\label{eq:pzr}
\end{equation}
$S(m,\eta,r)$ is the selection function, and $\mu(r) \equiv 5 \log r$
is the distance modulus. 
Note the typographical error in equation~11 of Paper I;
equation~(\ref{eq:pmgetaz}) is correct.  In this Appendix, we 
derive analytic approximations to equations~(\ref{eq:pmgetaz}) and
(\ref{eq:petagmz}) using the method of steepest descent.
We first consider the simple case of no selection ($S = 1$) in
A.1, and then consider distance-independent selection functions (A.2).
The MAT sample selection function does have a distance dependence; we
treat this case in A.3. In A.4 we further refine the approximation and summarize results.

\subsection{The Case of No Selection}
\label{sec:noselection}

Equation~(\ref{eq:pzr}) gives the probability that an object at
distance $r$ exhibits redshift $cz.$ 
That probability is greatest for $r=w,$ where $w$
is the ``crossing point'' defined implicitly by $cz = w + u(w)$. 
%(cf.\ \S~2.3).
%[MS: I've repeated the equation for clarity.  The alternative is to
%make it a displayed equation in 2.3, and refer to the equation number
%here.]
Expanding about the crossing point gives
$[cz - (r+u(r))] \approx -(r-w)(1+\uprime),$ where
$\uprime$ is the radial peculiar velocity derivative at the
crossing point. To the same order of approximation we
may write $(r-w)\approx w\ln(r/w).$  With these approximations,
equation~(\ref{eq:pzr}) becomes: 
%Combining these two
%approximations we may reexpress equation~9 of Paper I as follows:
\begin{equation}
P(cz|r) = \frac{1}{\sqrt{2\pi}\sigma_v}\,\exp\left(-\frac{\ln(r/w)^2}
{2\Delta_v^2}\right)\,,
\label{eq:pczapprox}
\end{equation}
where $\Delta_v\equiv\sigma_v/[w(1+u^\prime)].$ 

This approximation is valid under certain conditions:
First, 
there must be a unique crossing point $w.$
Second, $u(r)$ must be adequately linear within a few times $\sigma_v$
of $w.$  Third, $w$ must be sufficiently large 
%in comparison with $\sigv$
that the approximation $(r-w)/w\approx \ln(r/w)$ is a good one for $r$
within a few times $\sigv$ of $w$.
%[MS: Is that what you mean?]
%The first condition is absolute; if there are multiple
%crossing points the likelihood cannot be approximated and the
%full numerical integrations must be used. 
The second and third conditions are satisfied when
$\Delta_v \ll 1;$ in practice we found that $\Delta_v\leq 0.2$ was
usually sufficient to ensure good accuracy (after the
refinements discussed in \S~A.4).  
%This condition may fail,
%even for $\sigv/w \ll 1,$ when $\uprime \rightarrow -1$, % This signifies
%a ``flat'' zone in the redshift-distance relation. 

\def\deltf{\Delta_{{\rm TF}}}
We consider first the forward TF likelihood,
equation~(\ref{eq:pmgetaz}), in the case of no sample selection ($S = 1$).
Substituting
equation~(\ref{eq:pczapprox}) into equation~(\ref{eq:pmgetaz}) gives 
\begin{equation}
P(m|\eta,cz) = \frac{\int_0^{\infty}dr\, r^2 n(r)
\exp\left(-\frac{\ln(r/w)^2}
{2\Delta_v^2}\right)\exp\left(-\frac{\ln(r/d)^2}
{2\deltf^2}\right)}
{\sqrt{2\pi}\sigtf \int_0^{\infty}dr\,r^2 n(r)
\exp\left(-\frac{\ln(r/w)^2}
{2\Delta_v^2}\right)}
\label{eq:pmreduced}
\end{equation}
where 
$\deltf\equiv\frac{\ln10}{5}\sigtf$ and $d$ is the forward
TF distance (\S~2.3). 
The integrals in equation~\ref{eq:pmreduced} can be evaluated
analytically if we assume that the density field behaves
locally as a power law,
\begin{equation}
n(r) = n(w)\left(\frac{r}{w}\right)^{\gamma}\,.
\label{eq:powerdens}
\end{equation}
In practice, $n(r)$ is
not a true power law and the exponent is evaluated as
$\gamma(w) = \left[d\ln n(r)/d\ln r\right]_{r=w}.$  With this assumption,
equation~\ref{eq:pmreduced} may be written 
\begin{equation}
P(m|\eta,cz) = \frac{\int_{-\infty}^{\infty} e^{(3+\gamma)x} e^{-\frac{x^2}{2\Delta_v^2}}
e^{-\frac{(x-y)^2}{2\deltf^2}}dx}
{\sqrt{2\pi}\sigtf\int_{-\infty}^{\infty} e^{(3+\gamma)x} e^{-\frac{x^2}{2\Delta_v^2}}dx}\,.
\label{eq:pmred1}
\end{equation}
where $x\equiv \ln (r/w)$ and $y\equiv \ln (d/w).$ The numerator and
denominator integrals of equation~(\ref{eq:pmred1}) 
may be straightforwardly evaluated to obtain
\begin{equation}
P(m|\eta,cz) = \frac{\ln 10}{5}
\frac{1}{\sqrt{2\pi}\,\Delta_e}\,\exp\left\{-\frac{1}{2\Delta_e^2}
\left(y-(3+\gamma)\Delta_v^2\right)^2\right\} \,,
\label{eq:noselect}
\end{equation}
where
\begin{equation}
\Delta_e \equiv \left[\deltf^2
+\Delta_v^2\right]^{1/2}\,.
\label{eq:sigtf_eff}
\end{equation}

Equation~(\ref{eq:noselect}) has a simple interpretation.
When sample selection is neglected, 
the TF distance $d$ is log-normally distributed; the
expectation value of $\ln d$ is $\ln w + (3+\gamma)\Delta_v^2.$  
The fact that $E(\ln d)\neq \ln w$
is due to the Malmquist bias associated with velocity noise; there
is both a homogeneous ($3\Delta_v^2$) and an inhomogeneous ($\gamma\Delta_v^2$)
term.  Unlike the Malmquist bias
in a Method I approach which scales as $\deltf^2$ (cf.\ SW), the
bias here is proportional to $\Delta_v^2 \propto (\sigv/w)^2,$ 
which is generally much
smaller, and which decreases with distance. 

The expression for the inverse probability,
equation~(\ref{eq:petagmz}), is complicated by the presence of the
luminosity function $\Phi$ in both numerator and denominator. 
However, like the density field, this function varies
slowly on the scale relevant to the integration.
Consequently, we may treat it too as a power law for $r$ near $w:$
\begin{equation}
\Phi(m-\mu(r)) \approx \Phi(m-\mu(w))\left(\frac{r}{w}\right)^\lambda\,.
\label{eq:deflambda}
\end{equation}
Again, we evaluate the power-law exponent 
according to $\lambda(w)\equiv \left\{d\ln [\Phi(m-\mu(r))]/d\ln r\right\}_{r=w}.$
Once this is done, 
the integrals simplify in the same way
as for the forward relation, and
we find after similar manipulations 
\def\deltei{\Delta_{e,{\rm inv}}}
\def\yinv{y_{{\rm inv}}}
\begin{equation}
P(\eta|m,cz)= \frac{\ln 10}{5} \frac{1}{e}
\frac{1}{\sqrt{2\pi}\,\deltei}\,\exp\left\{-\frac{1}{2\deltei^2}
\left(\yinv-(3+\gamma+\lambda)\Delta_v^2\right)^2\right\} \,.
\label{eq:noselectinv}
\end{equation}
Here %In equation~(\ref{eq:noselectinv}),
$\yinv\equiv\ln(\dinv/w),$ where $\dinv$ is the inverse TF distance
and $e$ is the inverse TF slope (\S~2.3).
The fractional inverse TF distance error 
is given by 
\begin{equation}
\deltei \equiv \left[\Delta_\eta^2+\Delta_v^2\right]^{1/2}
\label{eq:defdelinv}
\end{equation}
where $\Delta_\eta\equiv(\ln 10/5) \sigeta/e.$

Comparison of Eqs.~\ref{eq:noselect} and~\ref{eq:noselectinv} reveals the close
analogy between the forward and inverse probability expressions when selection
is neglected.  
%Aside from the factor $e^{-1}$ out in front and
%the $\lambda$ in the exponent, one simply substitutes $\deltei$
%for $\Delta_e$ and
%and $\dinv$ for $d$ to obain the inverse from the forward probability. 
Such an analogy must indeed hold, for 
the two forms
of the TF relation contain the same information. 
The factor $e^{-1}$ in equation~\ref{eq:noselectinv}
simply renormalizes the probability density to $\eta$-space,
while the $\lambda$ reflects the luminosity function dependence of
the inverse expression. 

\subsection{The role of selection}
\label{sec:selection-role}

%By and large sample selection enters in a manner almost perfectly analogous
%to pure Method II.
%[MS: I've taken this sentence out because although I sort of know
%what it means (and I could educate myself by reading the appropriate
%section of SW); 98% of our readers will not understand it.)
In this section, we assume that the sample selection function has 
%We begin with the simplest case: forward TF relation, and the sample
%selection function has 
no {\em explicit\/} $r$-dependence, i.e., $S=S(m,\eta).$  We assume
the sample to be selected on a quantity $\xi$ with limiting value
$\xi_\ell$,  which is linearly related to
the TF observables: 
\begin{equation}
\xi(m,\eta) = a_1 -b_1 m - c_1\eta\qquad \hbox{with scatter $\sigma_\xi$}\,.
\label{eq:ximeta}
\end{equation}
%The fit is done {\it a priori}; see Willick [JW] I found your wording confusing here.
%\etal\ (1996) for the results for the Mark III samples. 
The quantities
$a_1,$ $b_1,$ and $c_1$ and $\sigxi$ were determined empirically for
the Mark III samples by Willick \etal\ (1995, 1996).
%We assume thoughout that the selection function has its
%``one-catalog'' form (cf.\ Willick 1994),
%In this case, 
Willick (1994) shows that: 
\begin{equation}
S(m,\eta) = \frac{1}{2} \left[1+\erf(\Axi(m,\eta)\right]\,,
\label{eq:select}
\end{equation}
where 
\begin{equation}
\Axi(m,\eta) \equiv \frac{\xi(m,\eta)-\xlim}{\sqrt{2}\sigxi}\,.
\label{eq:defaxiprime}
\end{equation}
%$\xi$ is the quantity on which selection is based (e.g., log diameter),
%$\xi(m,\eta)$ is its expected value given the TF observables,
%$\xlim$ its limiting value, and $\sigxi$ the scatter of observed
%about expected values. We further assume
%that the selection quantity is linearly related to the TF observables,
%\begin{equation}
%\xi(m,\eta) = a_1 -b_1 m - c_1\eta\,.
%\label{eq:ximeta}
%\end{equation}
%The coefficients $a_1,$ $b_1,$ and $c_1$ are previously determined constants (in 
%the case of the Mark III samples these constants were determined by
%Willick \etal\ 1996).

We define a TF-predicted apparent magnitude 
$m_r\equiv M(\eta)+5\log r$.  
%the corresponding predicted selection parameter is 
%\begin{equation}
%\Axi(m_r,\eta) = \frac{\xi(m_r,\eta)-\xlim}{\sqrt{2}\sigxi}\,.
%\label{eq:defAxi}
%\end{equation}
%[MS: This equation is completely redundant with the previous
%equation.] 
Then, using the identities derived by Willick (1994), 
%and the [JW] The next step has nothing to do with the approximations!
%approximations made in the previous subsection, 
the forward likelihood becomes: 
\begin{equation}
P(m|\eta,cz) = \frac{\left[1+\erf(\Axi(m,\eta))\right] }{\sqrt{2\pi}\sigtf}
\frac{\int r^2 n(r) P(cz|r) \exp\left(-\frac{\ln(r/d)^2}
{2\deltf^2}\right)\,dr }
{\int r^2 n(r) P(cz|r)
\left[1+\erf\left(\Axi(m_r,\eta)/\sqrt{1+\beta^2}\right)\right]\,dr}\,,
\label{eq:pfsel}
\end{equation}
where $\beta\equiv b_1\sigtf/\sigxi.$ 

The integral over $m$ has caused the ${\cal A}_\xi$
term in the deonominator to acquire
%deriving from the selection function in the denominator has acquired 
an $r$-dependence, although it did not start out with one. 
This complication makes it inconvenient to follow our previous procedure
exactly. Instead, we 
%The approach we will instead take, which is correct
%to the same order of approximation,
%is to 
treat this term % involving the error function in the denominator as
as constant across the effective range of integration, and take it
outside the integral; this is correct to the same order of
approximation.  This leaves us with a ratio of integrals we
have already evaluated. We then {\em require\/} that the resultant
probability density $P(m|\eta,cz)$ be properly normalized, yielding:
\begin{equation}
P(m|\eta,cz) = \frac{\ln 10}{5}\frac{1+\erf\left[\Axi(m,\eta)\right]}
{1+\erf\left[\frac{\Axi(m_0,\eta)}{\sqrt{1+\beta^2}}\right]}
\frac{1}{\sqrt{2\pi}\,\Delta_e}
\exp\left\{-\frac{1}{2\Delta_e^2}
\left(y-(3+\gamma)\Delta_v^2\right)^2\right\}\,,
\label{eq:select_forw}
\end{equation}
where
\begin{equation}
m_0 (\eta,w) \equiv  M(\eta)+5\log w + \frac{5}{\ln10}\left[3+\gamma\right]\Delta_v^2
\end{equation}
and
\begin{equation}
\beta\equiv \frac{b_1\sigma_e}{\sigxi}\qquad {\rm where\ }
\sigma_e\equiv 5\Delta_e/\ln 10\,.
\end{equation}
The effect of selection appears purely {\em outside\/} the
exponent now. Indeed, the role of selection is very similar
to what it was in pure Method II (Willick 1994), with a slightly
different evaluation of $\Axi$ and $\beta$ in the denominator.
%{\bf Can we give the reader an intuition as to what $m_0$ means?  It's
%a slightly bias apparent magnitude the galaxy would have had if it
%were at the distance implied by its $\eta$...}. 

The corresponding expression for the inverse relation follows
directly, given the analogy we drew between the two expressions in the
previous subsection:
\begin{equation}
P(\eta|m,cz) = \frac{\ln 10}{5}\,e^{-1} \frac{1+\erf\left[\Axi(m,\eta)\right]}
{1+\erf\left[\frac{\Axi(m,\eta_0)}{\sqrt{1+\beta^2}}\right]}
\frac{1}{\sqrt{2\pi}\,\Delta_e}
\exp\left\{-\frac{1}{2\Delta_e^2}
\left(y-(3+\gamma+\lambda)\Delta_v^2\right)^2\right\}\,,
\label{eq:select_inv}
\end{equation}
where 
\begin{equation}
\eta_0(m,w) \equiv \eta^0\left(m-\left[5\log w + \frac{5}{\ln 10}(3+\gamma+\lambda)\Delta_v^2\right]\right)
\end{equation}
and 
\begin{equation}
\beta \equiv \frac{c_1\sigma_{\eta,e}}{\sigxi}\qquad {\rm where\ } \sigma_{\eta,e}\equiv 
\sqrt{\sigma_\eta^2 + \left(5e/\ln 10 \Delta_v\right)^2}\,.
\end{equation}
Note the different
definition of $\beta$ in the inverse and forward cases.
In particular, if selection is $\eta$-independent ($c_1=0$), the
terms involving the error functions 
cancel, and $P(\eta|m,cz)$ reduces to the no selection case, as
expected. 
%{\bf No it doesn't!  $\eta \neq \eta_0$!  Notice also the
%awkward notation with the subscript and superscript 0's on $\eta$;
%this starts looking like GR!}

\subsection{Treating an explicitly distance-dependent selection function}
\label{sec:S(r)}

%Under certain circumstances (cf.\ Strauss \& Willick 1995) 
%the selection function acquires an explicit distance dependence,
%and must be written $S(m,\eta,r).$ 
If the selection function $S$ has an explicit distance dependence,
things get a bit more complicated.  In Willick \etal\ (1996), the data
for all the Mark III samples was fit to the form, 
\begin{equation}
\xi = \xi(m,\eta,r) = a_1 - b_1 m -c_1\eta -d_1 \log r\,;
\label{eq:ximetar}
\end{equation}
only MAT had a significantly non-zero value of $d_1$. 
However, $c_1=0$ for MAT; selection for MAT has no explicit $\eta$-dependence
and we take this into account in what follows.
Corresponding to equation~(\ref{eq:ximetar}) is an $r$-dependent $\Axi$ parameter,
\begin{equation}
\Axi(m,r) \equiv \frac{\xi(m,r)-\xlim}{\sqrt{2}\sigxi}\,,
\label{eq:aximetar}
\end{equation}
and thus the selection function $S(m,r)=\left[1+\erf(\Axi(m,r)\right]/2.$
%{\bf I wonder if this entire sentence can be dropped?  The notation is
%quite self-explanatory.} 

%Including the $r$-dependence of $S$ in the likelihood expressions
%requires the introduction of  
The main effect of distance-dependent selection 
is to introduce 
a new power-law exponent, $\alpha,$
into our earlier expressions, where
\begin{equation}
\alpha(m,w) \equiv \left.\frac{d\ln S}{d\ln r}\right|_{r=w}  
=  - \sqrt{\frac{2}{\pi}}\frac{d_1}{\ln 10\,\sigxi} \frac{e^{-\Axi^2}}
{\left[1+\erf(\Axi)\right]} \,.
\label{eq:alpha2}
\end{equation}
For the inverse relation, the addition of $\alpha$ is {\em all\/}
that is required to correct our expressions. Specifically,
\begin{equation}
P(\eta|m,cz) = \frac{\ln 10}{5} \frac{1}{e}
\frac{1}{\sqrt{2\pi}\,\Delta_{e,inv}}\,\exp\left\{-\frac{1}{2\Delta_{e,inv}^2}
\left(y_{inv}-(3+\gamma+\lambda+\alpha)\Delta_v^2\right)^2\right\} \,,
\label{eq:selectinv}
\end{equation}
There are no selection functions out in front because for MAT,
selection is $\eta$-independent. 
%Note, however, that the fact that $\alpha\neq 0$ means
%that the inverse TF probability {\em is\/} affected by selection.

For the forward relation, the fact that $\alpha$ depends on $m$
ruins the pure Gaussianity of the exponent.  Using the same approach
as we did
to derive equation~(\ref{eq:select_forw}), we assume that 
$\alpha$ varies slowly with $m$, take ${\cal A}_\xi$ out of the
denominator integral, and normalize after the fact.   
% one can readily derive a normalized probability. 
After some algebra, one finds
\begin{equation}
P(m|\eta,cz)\simeq \frac{\ln 10}{5}\,\frac{1+\erf\left(\Axi(m,w)\right)}
{1+\erf\left[\frac{\Axi(\overline{m},w)}{\sqrt{1+\beta^2}}\right]}
\,\frac{1}{\sqrt{2\pi}\,\Delta_e^\prime}\,\exp\left\{-\frac{1}{2\Delta_e^2}
\left(y-(3+\gamma+\alpha(m,w))\Delta_v^2\right)^2\right\} \,.
\label{eq:fselect_d}
\end{equation}
%{\bf Is there a prime in the $\delta$ inside the exponential?} JW - no
In equation~\ref{eq:fselect_d}, the individual terms have the following definitions:
\begin{equation}
\overline{m} \equiv m_0+\alpha(m_0,w)\frac{5}{\ln 10}\Delta_v^2\,,
\end{equation}

\begin{equation}
\Delta_e^\prime \equiv\Delta_e\left[1-{{5}\over{\ln 10}}\Delta_v^2
{{d\alpha}\over {dm}}\right]^{-1} = 
\Delta_e\left[1-{{5\sqrt{2}\, b_1 \alpha(m,w) \Delta_v^2}\over{\ln 10\,\sigma_e}}\left(
\Axi - {{\alpha(m,w) \ln 10\, \sigma_\xi}\over {\sqrt{2}\,d_1}}\right)\right]^{-1}
\end{equation}

\begin{equation}
\beta \equiv \frac{b_1\sigma_e^\prime}{\sigxi}\qquad {\rm where\ }
\sigma_e^\prime \equiv {{5\Delta_e^\prime}\over{\ln 10}}\,. 
\end{equation}
%It is the $m$-dependence of the $\alpha$ in the exponent of
%equation~(\ref{eq:fselect_d}) that requires small change
%from $\Delta_e$ to $\Delta_e^\prime$ in the normalization (but
%not in the exponent itself).

%Finally, we need an expression for $d\alpha/dm.$ This turns
%out to be, after some algebra,
%\begin{equation}
%\frac{d\alpha}{d m} = \frac{\sqrt{2}\beta}{\sigma_e^\prime}\,\alpha\,
%\left[A - \frac{\alpha}{f\sqrt{\pi}}\right]\,,
%\label{eq:dadm}
%\end{equation}
%where $f\equiv \sqrt{2/\pi}\,d_1/(\ln 10\,\sigxi)$ is the coefficient
%that appeared above in the definition of $\alpha$ and the other
%terms are as defined above. In equation~\ref{eq:dadm}, all quantities
%depending on magnitude are to be evaluated at $m_0$ as
%defined above. 
%[MS: I just did the direct substitution above, to save some room.]

\subsection{Final Refinement and Summary}

Our original approximation to $P(cz|r)$, equation~(\ref{eq:pczapprox}), was correct
to first order in $\sigv.$ This leads to systematic 
inaccuracies in two regimes: small distances ($\simlt 2000\ \kms$),
where the approximation $(r-w)/w\approx \ln (r/w)$ loses 
accuracy, and in regions of velocity field curvature, when
$\udoubleprime\sigv$ is comparable to $\uprime.$ 
We extend its regime of validity 
by making second-order corrections for
these effects. 

To second order in $x \equiv \ln (r/w)$ we find
%$(r-w)/w = \ln(r/w)+\left[\ln(r/w)\right]^2/2.$ 
$(r-w)/w = x + x^2/2$. Using
this and a second order Taylor expansion of $u(r)$ about
$w,$ and retaining only terms of order $(\sigv/w)^2$
in the exponent, we find after some algebra
\begin{equation}
\exp\left[-\frac{1}{2\sigv^2}\left(cz-[r+u(r)]\right)^2\right] =
e^{-x^2/2\Delta_v^2}\times f(r)\,,
\label{eq:pcznew}
\end{equation}
where %$x=\ln r/w$ and
\begin{equation}
f(r) \equiv e^{(1+\varepsilon)\,x^3/2\Delta_v^2}\ , \ 
\varepsilon \equiv \frac{\udoubleprime w}{1+\uprime}\,,
\label{eq:defeps}
\end{equation}
where $\uprime$ and $\udoubleprime$ are evaluated at $w.$
The term $e^{-x^2/2\Delta_v^2}$ in
equation~(\ref{eq:pcznew}) is just our original
approximation for $P(cz|r),$ equation~(\ref{eq:pczapprox}), and $f(r)$
%The term which multiplies it---let us call it $f(r)$---
is the second-order correction.
It is non-Gaussian in $\ln(r/w)$ and thus cannot be analytically
integrated as before. 

We thus treat it as we have other slowly-varying terms:
% such as density and luminosity function: 
we approximate it as a power
law in the vicinity of the crossing point. However, because of its
cubic nature, the local logarithmic derivative is identically zero. 
%were we
%to calculate the power-law exponent, $\nu,$  by 
%direct logarithmic diffentiation we would find
%$\nu=\left[d\ln f/d\ln r\right]_{r=w}=0.$ This results from
%the cubic dependence of the exponential. 
We thus proceed heuristically by calculating the power-law exponent as
a finite difference over an interval of $\ln r$ of $\pm g \Delta_v$, where
$g$ is of order unity: 
%To get around
%this problem we proceed heuristically. Rather than
%as a derivative, we compute the power law exponent 
%as a finite difference over the appropriate interval of $\ln r,$
%which must be of order $\Delta_v.$
%Thus,
%\begin{equation}
%\nu \sim \frac{\ln f(r=\Delta_v) - \ln f(r=-\Delta_v)}{2\Delta_v}
%= -\frac{1}{2}\left(1+\varepsilon\right)
%\label{eq:approxnu}
%\end{equation}
\begin{equation}
\nu \sim \frac{\ln f(x=g\Delta_v) - \ln f(x=-g\Delta_v)}{2\,g\Delta_v}
= -\frac{g^2}{2}\left(1+\varepsilon\right)
\label{eq:approxnu}
\end{equation}
We calibrated the appropriate value of $g$ by varying it until 
we maximized agreement between the exact and approximate
likelihoods.  This happened at $g = 1.5,$ and thus
the correct exponent is
\begin{equation}
\nu=-1.1\left(1+\varepsilon\right)\,.
\label{eq:finalnu}
\end{equation}

This leads to the final forms of the analytic approximation to the
\velmod\ likelihoods. For the forward relation, $P(m|\eta,cz)$
is given by equation~(\ref{eq:select_forw}) for A82, W91PP, and CF
(the samples for which the selection function has no explicit
distance dependence) and 
by equation~(\ref{eq:fselect_d}) for MAT. For the inverse
relation, $P(\eta|m,cz)$ is given by equation~(\ref{eq:select_inv})
for A82, W91PP, and CF and by equation~(\ref{eq:selectinv}) for MAT.
However, in all of these equations, the quantity 
$3+\gamma$ is replaced by $3+\gamma+\nu,$ where
$\nu$ is given by equation~(\ref{eq:finalnu}), and
$\varepsilon$ is given by equation~(\ref{eq:defeps}).
Note that the definition of $\nu$ is such that the
homogeneous Malmquist bias term is reduced from
$3\Delta_v^2$ to $1.9\Delta_v^2.$ This is a significant
effect for distances $\simlt 2000\ \kms,$ and thus
the refinement discussed here is crucial for extending
the regime of validity of the approximation to small distances.

\end{document}